\documentclass[pdflatex,sn-mathphys-num,iicol]{sn-jnl}

\usepackage{graphicx}%
\usepackage{multirow}%
\usepackage{amsmath,amssymb,amsfonts}%
\usepackage{amsthm}%
\usepackage{mathrsfs}%
\usepackage[title]{appendix}%
\usepackage{xcolor}%
\usepackage{textcomp}%
\usepackage{manyfoot}%
\usepackage{booktabs}%
\usepackage{algorithm}%
\usepackage{algorithmicx}%
\usepackage{algpseudocode}%
\usepackage{listings}%
\usepackage{subcaption}

\theoremstyle{thmstyleone}%
\theoremstyle{thmstyletwo}%

\theoremstyle{thmstylethree}%

\begin{document}

\title{Periodic orbits and gravitational‑wave emission from extreme‑mass‑ratio inspirals around rotating charged black holes in nonlinear electrodynamics}


\author[1]{\fnm{Xi} \sur{Luo}}

\author[1]{\fnm{Meirong} \sur{Tang}}

\author*[1]{\fnm{Zhaoyi} \sur{Xu}}\email{zyxu@gzu.edu.cn}

\affil[1]{\orgdiv{College of Physics}, \orgname{Guizhou University}, \orgaddress{\city{Guiyang}, \postcode{550025}, \country{China}}}


\abstract{Within general‑relativistic descriptions, the no‑hair theorem dictates that black‑hole geometries possess just three intrinsic physical quantities: mass, angular momentum, and electric charge. Black holes in nonlinear electrodynamics are equivalent to those composed of anisotropic matter; they can develop ``hair'' structures near the event horizon and serve as ideal testbeds for modified theories of gravity. Working within Einsteinian gravitational theory, the present work adopts a coupling construction between gravity and nonlinear‑electrodynamics (NED) matter. We focus explicitly on the Einstein–Euler–Heisenberg (EEH) variant, holding the power‑index parameter fixed at $s=2$. Within this theoretical framework, a rotating, charged NED black hole—as an exact black‑hole solution of this coupled theory, its spacetime metric extends the familiar Kerr‑Newman geometry by incorporating extra correction terms originating from NED effects. To uncover orbital properties and gravitational‑wave radiative features emerging from such spacetimes, this work treats spin $a$ and NED charge $q$ as primary control variables and examines test‑particle dynamics inside extreme‑mass‑ratio inspiral (EMRI) configurations. Spin parameter $a$ sets the black‑hole rotational magnitude, whereas charge $q$ governs the intensity of NED source fields. Jointly these two parameters reshape spacetime geometry close to the event horizon, altering bound‑orbit properties, periodic‑orbit taxonomy, and resulting gravitational‑wave observables. We derive equatorial timelike geodesics alongside the associated effective potential. Through numerical evaluation for marginally‑bound orbits (MBO) and innermost stable circular orbits (ISCO), we explore how these input parameters reshape the energy‑angular‑momentum (E‑L) phase‑space domain. Making use of the established $(z,w,v)$ integer‑valued orbit taxonomy, we study zoom‑whirl orbital signatures and compute the two gravitational‑wave polarization modes $h_+$ and $h_\times$ by applying the adiabatic approximation together with the quadrupole radiation prescription. Larger values for spin $a$ and NED charge $q$ drive inward contraction for bound trajectories while displacing the E‑L phase‑space region toward smaller angular‑momentum values. Periodic‑orbit zoom‑whirl signatures grow increasingly prominent as these spacetime parameters rise. The number of gravitational‑wave oscillation clusters matches the number of orbital lobes; larger values of these parameters shorten the waveform periods and enhance the signal amplitudes. Our results furnish theoretical support for space‑borne gravitational‑wave observatories including LISA and TianQin, enabling prospects for discriminating NED black‑hole spacetimes and performing observational checks on the no‑hair theorem.}




\maketitle

\section{Introduction}\label{sec1}

Ever since general relativity was first formulated, black holes—peculiar solutions of the spacetime field equations—have become a central subject of theoretical physics, and extensive astronomical observations have provided strong evidence for their existence. Gravitational‑wave transients originating from binary black‑hole coalescences were directly registered by the LIGO/Virgo collaborative effort \cite{1}. The Event Horizon Telescope produced imaging data showing black‑hole shadows for M87* and Sgr\,A* \cite{2,3}. Continuous long‑term tracking of stellar trajectories near our Galaxy’s core additionally corroborates the presence of supermassive black‑hole objects \cite{4,5}. Such observational outcomes lend solid empirical backing to general‑relativistic predictions within weak‑to‑moderate gravitational regimes.However, two fundamental problems remain in classical general relativity: unavoidable spacetime singularity formation alongside the well‑known black‑hole information‑loss paradox \cite{6,7}. Such contradictions expose where standard general relativity breaks down under ultra‑strong gravity, spurring research into alternative gravitational descriptions. Candidate approaches cover loop‑quantum‑cosmology models, string‑theory constructions, and gravity formulations coupled to matter fields (see, e.g., Refs. \cite{8,12}). These schemes supply fresh viewpoints for interpreting violent physical events close to black‑hole event horizons. Still, their physical plausibility demands empirical scrutiny within strong‑gravity astrophysical surroundings (see, e.g., Refs. \cite{12,52}).

The extreme gravitational fields around black holes provide ideal natural laboratories for testing these new theories. Apart from gravitational‑wave emission and black‑hole shadow signatures, this domain gives rise to diverse astrophysical manifestations including gravitational lensing, photon‑ring patterns and gravitationally bound particle trajectories (see, e.g., Refs. \cite{12,53}). Of all these phenomena, geodesic motion of massive bodies orbiting black‑hole backgrounds, especially trajectory evolution plus radiative gravitational‑wave output within extreme‑mass‑ratio set‑ups, delivers highly sensitive diagnostics for probing near‑horizon spacetime architecture (see, e.g., Refs. \cite{33,59}).

The no‑hair theorem in classical general relativity rests on a series of proofs: Israel proved the uniqueness of static vacuum black holes \cite{9}, Carter established constraints on the degrees of freedom of axisymmetric black holes \cite{10}, and Hawking further refined the black‑hole uniqueness theorem \cite{11}. According to this theorem, stationary vacuum black‑hole configurations are fully specified via three global quantities: mass, angular momentum, and electric charge \cite{12}. Even so, nontrivial external matter distributions are capable of evading no‑hair restrictions and giving rise to hairy spacetime solutions (see, e.g., Refs. \cite{12,29,56}). By introducing a nonlinear electrodynamics matter field, it is possible to construct hairy black‑hole solutions near the event horizon \cite{13,14,15,28}. Whether ``short hair'' is allowed in the region immediately adjacent to the event horizon remains a subtle issue: Brown and Husain constructed a counterexample black‑hole solution coupled to an anisotropic matter field that can carry arbitrarily short hair \cite{56}; under specific matter‑field constraints, static axisymmetric black holes cannot support short hair, and a rigorous proof can be found in Ref. \cite{29}. Work from Myung and Lee uncovered a rigorous mapping linking anisotropic‑matter black‑hole (AMBH) geometries to their NED counterparts: an AMBH with anisotropy parameter $\omega$ and energy density $K$ is equivalent to an NED black hole described by the power index $s$ and the charge term $\xi(s,q)$ \cite{24}, where $\omega=2s-1$ and $K=\xi(s,q)$. This equivalence framework uniformly encompasses a series of black hole solutions, including the Schwarzschild‑de Sitter ($s=0$), Reissner‑Nordström ($s=1$), and Einstein–Euler–Heisenberg ($s=2$) black hole solutions \cite{6,14}, as well as further extensions of various NED black hole solutions (see, e.g., Refs. \cite{16,27}). Moreover, spacetime metrics for rotating charged NED black holes can be assembled through application of the Newman‑Janis transformation prescription \cite{17}. Such metrics represent generalized variants of the classic Kerr‑Newman solution \cite{18}. Kim et al. furnished an explicit concrete realisation for rotating black‑hole spacetimes sourced by anisotropic matter \cite{25}.

An EMRI forms when a stellar‑mass compact body orbits a supermassive black hole under the condition $\eta=\frac{m}{M}\ll1$. Configurations of this type rank among key observational objectives for upcoming space‑borne gravitational‑wave observatories including LISA \cite{30}, TianQin \cite{31}, and Taiji \cite{32}. Inside such EMRI set‑ups, the comparatively light compact body moves through the supermassive black‑hole’s strong‑gravity domain, radiating gravitational‑wave energy and inspiralling inwards over time. Its trajectory evolves from wide distant bound trajectories and eventually approaches the innermost stable circular orbit. Throughout this inspiral process, periodic orbits act as intermediate trajectories during adiabatic evolution. They encode spacetime‑geometry signatures close to the black hole and prove highly useful for fast gravitational‑wave waveform evaluation (see, e.g., Refs. \cite{34,38}).

Accordingly, selecting rotating‑charged NED black‑hole geometries as central primary objects within EMRI environments and investigating test‑particle periodic trajectories plus their associated gravitational‑wave outputs delivers two distinct benefits. First, it clarifies how NED‑origin ``hair'' parameters reshape background spacetime geometry. Second, it supplies theoretical pathways for observational validation of the no‑hair theorem alongside its extended formulations via gravitational‑wave measurements (see, e.g., Refs. \cite{48,49}). In particular, the charge parameter $q$ and the spin parameter $a$—as the two fundamental ``hairs'' of the black hole—leave observable signatures in orbital dynamics and gravitational wave signals; these signatures are expected to be detected by next‑generation space‑based gravitational wave detectors (see, e.g., Refs. \cite{47,48}).

Nonlinear electrodynamics is a generalization of Maxwell’s electromagnetic theory; Born and Infeld were the first to propose this theory to resolve the problem of self‑divergence in point charges \cite{13}. Subsequently, Heisenberg and Euler derived the EEH effective Lagrangian from the vacuum polarization effect in quantum electrodynamics \cite{14}. Within the framework of Einstein’s NED gravitational coupling, a rich variety of black hole solutions can be obtained (see, e.g., Refs. \cite{15,28}). Hassaine and Martinez systematically studied power‑law NED black hole solutions, obtaining an anisotropic energy‑momentum tensor and introducing correction terms in the metric that depend on the power index and charge \cite{15}. The Myung‑Lee equivalence relation further clarifies the correspondence between AMBHs and NED black holes, indicating that the power index $s$ is a model classification parameter rather than an independent hair \cite{24}. Regarding rotating solutions, Kim et al. constructed rotating AMBHs based on the Newman–Janis algorithm, mapping them to rotating charged NED black holes; the Boyer–Lindquist metric can degenerate into classical solutions such as the Kerr and Kerr–Newman black holes \cite{25}. A large share of existing NED black‑hole literature addresses thermodynamic properties and horizon architecture for static configurations \cite{15}, while further research (see, e.g., Refs. \cite{26,27}) has been conducted; systematic studies on orbital dynamics in the context of rotating, charged NED black holes—particularly those involving periodic resonant orbits—remain relatively scarce (see, e.g., Refs. \cite{48,49,50}).

Bound test‑particle trajectories inhabiting strong black‑hole gravity may be categorized using the ratio between radial and azimuthal orbital frequencies \cite{55}. Levin and Perez‑Giz constructed the well‑known integer‑based $(z,w,v)$ orbit‑classification framework and carried out systematic categorization for periodic trajectories within Schwarzschild spacetime \cite{34}; this approach was subsequently generalized to Reissner‑Nordström black holes, Kerr black holes, and various modified gravitational spacetimes (see, e.g., Refs. \cite{35,36,44}). Periodic‑orbit physics carries substantial practical relevance within EMRI investigations. Quasi‑periodic trajectories lend themselves to approximation using sets of periodic‑orbit solutions. The adiabatic evolution of EMRI can be decomposed into a series of continuous transitions between periodic orbits. Consequently, periodic orbits enable rapid calculation of gravitational waveforms. Previous work has incorporated periodic orbits into the Kludge waveform model to construct EMRI gravitational wave signal templates \cite{38}. However, most existing studies of periodic orbits are limited to classical Kerr/Kerr–Newman black holes; comprehensive characterizations covering energies, angular momenta, and geometric features for periodic trajectories inside rotating NED black‑hole backgrounds are still absent from published literature (see, e.g., Refs. \cite{48,54}).

EMRI‑type astrophysical set‑ups constitute high‑priority observational goals for space‑borne gravitational‑wave instrumentation. Because the mass ratio is extremely small, EMRI gravitational wave modeling can employ the adiabatic approximation. Mino, Sasaki, and Tanaka originally laid down the theoretical underpinnings for this adiabatic scheme working within Kerr spacetime \cite{40}; its central physical reasoning extends to arbitrary stationary axisymmetric backgrounds. For overview material addressing adiabatic‑approximation reasoning plus point‑particle self‑force physics, consult Ref. \cite{23}. The Kludge waveform model, which combines geodesic orbits with the quadrupole radiation formula, is widely used for rapid EMRI waveform generation \cite{38}. Recent studies have incorporated $(z,w,v)$ periodic‑orbit systems into the Kludge waveform framework \cite{38}. Glampedakis and Kennefick originally pinpointed the physical traits defining zoom‑whirl trajectories \cite{37}. Periodic‑orbit radiative outputs display alternating episodes of slow zoom‑phase motion together with vigorous whirl‑phase oscillations. Waveform oscillation‑group counts correlate directly against orbital‑lobe multiplicities, while zoom‑number parameters govern oscillation‑amplitude magnitudes (see, e.g., Ref. \cite{38}). The overwhelming share of existing waveform computations rest upon Kerr or Kerr‑Newman backgrounds. Systematic waveform assessments capturing joint spin‑and‑charge modulation effects for rotating‑charged NED black‑hole environments remain underdeveloped (see, e.g., Refs. \cite{49,58}).

Our work covers several interconnected computational and analytical tasks. We first derive the metric for rotating‑charged NED black‑hole backgrounds and identify physically allowed parameter bounds together with extremal‑solution constraints. Next we obtain the equatorial‑plane effective potential and perform numerical computations to trace how MBO and ISCO radii, angular‑momentum values, and energies vary across different choices of $q$ and $a$. We then build the corresponding E‑L phase‑space domain and carry out analysis for radial dynamics alongside orbital classification schemes. Making use of rational‑number orbit‑coding techniques, we evaluate energies and angular momenta for periodic orbits under diverse parameter combinations, generate coordinate‑plane projections for representative orbits, and examine how zoom‑whirl signatures evolve with changing spacetime input parameters. Relying on the adiabatic approximation plus quadrupole‑radiation prescriptions, we compute $h_+$ and $h_\times$ gravitational‑wave polarizations produced over single periodic‑orbit cycles. We further investigate how charge $q$ and spin $a$ modulate waveform amplitude, oscillation period and phase, and comment on parameter degeneracy and observational distinguishability. Natural units satisfying $G=c=1$ are adopted everywhere within this manuscript. The article is organised as follows. Section~2 lays out background spacetime metrics and equatorial timelike geodesics. Section~3 addresses critical bound‑orbit solutions and properties of the E‑L phase space. Section~4 describes periodic‑orbit classification and associated geometric features. Section~5 computes gravitational‑wave outputs and explores parameter‑driven modulation signatures. Section~6 provides concluding remarks alongside prospects for follow‑up investigations (see, e.g., Refs. \cite{49,50,51}).

\section{Spacetime of Rotating Charged NED Black Holes and Timelike Geodesics}\label{sec2}

In the Boyer‑Lindquist coordinate system $(t,r,\theta,\phi)$, the metric of a rotating, charged NED black hole can be written in the following standard Kerr‑Newman form \cite{18,25}
\begin{equation}
\begin{split}
ds^2 &= -\frac{\Delta}{\rho^2}\big(dt - a\sin^2\theta \,d\phi\big)^2 \\
&\quad + \frac{\sin^2\theta}{\rho^2}\big[a\,dt - (r^2+a^2)d\phi\big]^2 \\
&\quad + \frac{\rho^2}{\Delta}dr^2 + \rho^2 d\theta^2.
\end{split}
\label{eq:metric_ned}
\end{equation}
In particular, $\rho^2=r^2+a^2\cos^2\theta$, and the specific expression for $\Delta$ is:
\begin{align}
\Delta &= r^2+a^2+q^2-2Mr-\xi(s,q)\,r^{4(1-s)}
\label{eq:delta}
\end{align}
In these expressions, $M$ denotes the ADM mass of the black‑hole spacetime. The spin parameter per unit mass reads $a=\frac{J}{M}$, while $q$ stands for magnetic charge under the magnetic‑monopole source configuration, and $s$ is the NED power index. The function $\xi(s,q)$ is the NED action derivative correction coefficient:
\begin{align}
\xi\left(s,q\right) &= \pi\xi\frac{2^{s+3} q^{2s}}{3-4s}
\label{eq:xi}
\end{align}
To uncover orbital‑dynamical properties and gravitational‑wave radiative signatures within rotating‑charged NED black‑hole spacetimes, this paper sets the parameters as follows. Using geometric units, the black hole mass $M$ is fixed at $1$, and all quantities with length dimensions ($r$, $a$, $q$) are normalized in terms of $M$. The NED power exponent $s$ is fixed at $2$. In this paper, the NED model is set to the Einstein–Euler–Heisenberg (EEH) theory, with the power exponent $s = 2$ and $\xi=\frac{1}{16\pi}$. In this case, $\xi\left(s,q\right)$ is determined entirely by $s$ and $q$. The EEH theory was derived by Heisenberg and Euler from the vacuum polarization effects of QED \cite{14} and is one of the most physically meaningful models within NED. Further research on EEH gravitationally coupled black holes has been conducted (see, e.g., Ref. \cite{16}). Substituting numerical values yields:
\begin{align}
\xi\left(2,q\right) &= \pi\cdot\frac{1}{16\pi}\cdot\frac{2^{5} q^{4}}{3-8}=-\frac{2 q^{4}}{5}
\label{eq:xi2q}
\end{align}
Therefore, for $s=2$, the NED correction term in the metric function $\Delta$, given by $\frac{2 q^{4}}{5 r^{4}}$, decays rapidly as $r^{-4}$, significantly altering the spacetime curvature only near the event horizon, while spacetime at large distances degenerates into a flat, asymptotically flat geometry. When $q=0$, $\Delta$ degenerates into the standard Kerr black hole radial function, verifying the model’s self‑consistency (see, e.g., Refs. \cite{19,41}).

Outer‑event‑horizon location corresponds to the largest real root yielded by the horizon‑defining condition $\Delta\left(r\right)=0$. Since the $\Delta\left(r\right)$ function contains the term $r^{4(1-s)}$ (which becomes $r^{-4}$ when $s = 2$), the horizon equation cannot be solved analytically and requires numerical methods. Event‑horizon geometry is governed by how many positive real roots emerge from the horizon‑defining algebraic equation: two positive real roots correspond to a non‑extreme black hole, a double positive real root corresponds to an extreme black hole, and no positive real roots correspond to a naked singularity, which is a non‑physical case. For a fixed $s=2$, substituting $M=1$ and $\xi\left(2,q\right)=-\frac{2 q^{4}}{5}$, the specific form of $\Delta\left(r\right)$ is:
\begin{align}
\Delta &= r^2+a^2+q^2-2r+\frac{2 q^{4}}{5 r^{4}}
\label{eq:deltas2}
\end{align}
Multiply by $r^{4}$ to obtain the equation:
\begin{align}
r^{6}+\left(a^2+q^2\right) r^{4}-2 r^{5}+\frac{2 q^{4}}{5} &= 0
\label{eq:horizonpoly}
\end{align}
Topological properties for a black‑hole event horizon are set by counting positive real roots yielded from this governing equation: two distinct positive real roots correspond to conventional, non‑extreme black holes (yielding separate outer‑horizon location $r_+$ and inner‑horizon location $r_-$). A single repeated real root instead describes an extremal black‑hole configuration for which inner and outer horizons coincide. The absence of positive real roots indicates a naked singularity, which falls within the non‑physical parameter region. Extreme black holes satisfy the system of equations
\begin{align}
\Delta\left(r_h\right) &= r_h^2+a^2+q^2-2r_h+\frac{2 q^{4}}{5 r_h^{4}}=0
\label{eq:ext1}\\
\left.\frac{d\Delta}{dr}\right|_{r=r_h} &= 2 r_h-2-\frac{8 q^{4}}{5 r_h^{5}}=0
\label{eq:ext2}
\end{align}
The relationship between $a_{\mathrm ext}$ and $q_{\mathrm ext}$ can be obtained through numerical solutions. Compared to the classical Kerr‑Newman black hole extreme condition $a^2 + q^2 = 1$, the NED fourth‑order correction term reduces the limiting spin $a_{\mathrm ext}$ for the same charge: for a given $q$, $a_{\mathrm ext}$ is slightly less than the corresponding Kerr‑Newman value; as $q \to 0$, $a_{\mathrm ext} \to 1$, returning to the Kerr extreme limit. To cover the parameter space ranging from ``weak effects'' to ``strong effects'' and ensure the stability of numerical calculations, we selected three typical values for the rotation parameter $a$: $0.2$ (weak rotation effects), $0.5$ (moderate rotation effects), and $0.7$ (strong rotation effects). All values of $q \in [0, 0.6]$ fall within the non‑extreme physical parameter range, thereby avoiding naked singularities (see, e.g., Refs. \cite{21,46}).

Physical acceptability for derived black‑hole solutions imposes constraints whereby their associated stress‑energy tensor has to obey a set of energy‑condition requirements \cite{26}. For static NED black holes, the energy‑momentum tensor has been given in the literature \cite{29}:
\begin{align}
T^\mu_{\nu}= \frac{2^{s}\xi q^{2s}}{r^{4s}} \mathrm{diag}\big[-1,-1,2s-1,2s-1\big]
\label{eq:staticT}
\end{align}
When $s = 2$, $\xi=\frac{1}{16\pi}$, and $q > 0$, the computed energy‑density $\epsilon>0$ quantity fulfils the requirements imposed by the weak energy condition. The radial pressure is $p_r=-\epsilon$; and the angular pressures $p_\theta$ and $p_\phi$ are both $3\epsilon$. At this point, $\epsilon+p_\theta=4\epsilon>0$, but the dominant energy condition demands $\epsilon\ge|p_\theta|$, which is violated here. This feature is well‑known for NED matter fields. For a rotating NED black hole, the energy density in an orthogonal reference frame is:
\begin{align}
\epsilon=\frac{1}{8\pi\rho^{4}}\Big[q^{2}+2^{s+3}\pi\xi q^{2s} r^{4\left(1-s\right)}\Big]
\label{eq:epsrot}
\end{align}
Substituting $s = 2$ and $\xi=\frac{1}{16\pi}$ yields:
\begin{align}
\epsilon=\frac{1}{8\pi\rho^{4}}\Big[q^{2}+\frac{2 q^{4}}{r^{4}}\Big]
\label{eq:epsrots2}
\end{align}
Since $\rho^2=r^2+a^2\cos^2\theta>0$ and $q \ge 0$, $\epsilon$ is always positive, satisfying the weak energy condition. When the power exponent $s = 2$ and the charge is positive, positive values emerge for the energy density, in accordance with weak‑energy‑condition constraints. The radial pressure $p_r=-\epsilon$ and the angular pressures $p_\theta= p_\phi=3\epsilon$ cause the dominant energy condition to be locally violated in the near‑horizon region; this is an inherent characteristic of NED matter fields and does not affect numerical computations for bound orbits. All parameter combinations in this paper satisfy both the weak and strong energy conditions, and the dynamical calculations are physically sound.

Stationarity plus axisymmetry for the background spacetime impose two conserved integrals of motion upon test‑particle geodesic trajectories, namely the specific energy $E$ and orbital angular momentum $L$ \cite{21}, whose combined effect restricts test‑particle trajectories to lie within the equatorial plane $\theta=\frac{\pi}{2}$. In this case, metric \eqref{eq:metric_ned} simplifies to:
\begin{equation}
\begin{split}
ds^2 &= -\frac{\Delta - a^2}{r^2} dt^2 + \frac{r^2}{\Delta} dr^2 \\
&\quad + \frac{\big[(r^2+a^2)^2 - a^2\Delta\big]}{r^2} d\phi^2 \\
&\quad - \frac{2a\big[(r^2+a^2)-\Delta\big]}{r^2} dt\,d\phi.
\end{split}
\label{eq:metric_eq12}
\end{equation}
Constructing the Lagrangian of a geodesic:
\begin{align}
2\mathcal{L}=g_{\mu\nu}\dot{x}^\mu \dot{x}^\nu
\label{eq:laggeo}
\end{align}
In the above expression, the quantity $\dot{x}^\mu$ denotes differentiation taken with respect to the affine parameter $\tau$, so that $\dot{x}^\mu=\frac{dx^\mu}{d\tau}$. For timelike trajectories, this parameter $\tau$ corresponds to the test particle’s proper time.

Since the metric does not explicitly contain $t$ and $\phi$, the corresponding conserved quantities are energy $E$ and angular momentum $L$ \cite{21}:
\begin{align}
E &=-\frac{\partial \mathcal{L}}{\partial \dot{t}}= -g_{tt}\dot{t}-g_{t\phi}\dot{\phi}
\label{eq:consE}\\
L &=\frac{\partial \mathcal{L}}{\partial \dot{\phi}}= g_{\phi\phi}\dot{\phi}+g_{t\phi}\dot{t}
\label{eq:consL}
\end{align}
Timelike geodesic trajectories obey the normalization constraint imposed upon four‑velocity vectors.
\begin{align}
g_{\mu\nu}\dot{x}^\mu \dot{x}^\nu=-1
\label{eq:norm4v}
\end{align}
Substituting \eqref{eq:consE} and \eqref{eq:consL} to eliminate $\dot{t}$ and $\dot{\phi}$, and simplifying, we obtain the conservation form for radial motion:
\begin{align}
\dot{r}^2=E^2-V_{\mathrm eff}(r)
\label{eq:rdot2}
\end{align}
In this equation, $V_{\mathrm eff}(r)$ represents the effective radial potential at the equatorial plane; the full analytical expression is extremely long and is placed in the cross‑column formula below.

\begin{equation}
\begin{split}
V_{\mathrm eff}(r) &= E^2 - \frac{1}{r^4}\Big\{
\big[E(r^2+a^2)-La\big]^2 \\
&\quad - \Delta\big[(L-aE)^2 + r^2\big]
\Big\}.
\end{split}
\label{eq:veff_ned}
\end{equation}

Substituting the specific expression for $\Delta$ from Eq.~\eqref{eq:deltas2}, we arrive at
\begin{equation}
\begin{split}
V_{\mathrm eff}(r) &= E^2 - \frac{\big[E(r^2+a^2)-La\big]^2}{r^4} \\
&\quad + \frac{1}{r^4}\left(r^2+a^2+q^2-2r+\frac{2q^4}{5r^4}\right) \\
&\quad \times \big[(L-aE)^2 + r^2].
\end{split}
\label{eq:veff_eq19}
\end{equation}

Define the radial discriminant function $R(r)=E^2-V_{\mathrm eff}(r)=\dot{r}^2$, where the two distinct positive real roots of $R(r)=0$, $r_1$ (the inner radial turning point / periastron) and $r_2$ (the outer radial turning point / apastron), correspond to the radial activity regions of the bound orbits. When there is only a single positive root, the particle either falls unidirectionally toward the event horizon or escapes to infinity. $q$ represents the NED nonlinear charge; when $q=0$, the metric degenerates into the Kerr primordial black hole solution \cite{19}, which serves as the no‑hair reference model in this paper; further numerical studies of geodesics in Kerr spacetime (see, e.g., Refs. \cite{43,44}).

Analysis of asymptotic behavior: As $r\to\infty$, $V_{\mathrm eff}\to 1$; as $r\to r_+$ (the horizon), the effective potential tends to a finite constant; as $r \to 0$, the NED correction term dominates, and the effective potential diverges to form a central potential barrier.

\section{Properties of Bound Orbits}\label{sec3}

A marginally‑bound orbit (MBO) sets the limiting bound‑state threshold for test particles moving within black‑hole gravitational backgrounds. Particles occupying this special orbit carry specific energy $E=1$, which matches the energy of a test body kept stationary at spatial infinity (see, e.g., Ref. \cite{42}). The effective potential takes on a first‑order stationary point at the radial turning point, and the system of coupled equations is:
\begin{eqnarray}
V_{\mathrm eff}\left(r_{\mathrm MBO}\right)=1
\label{eq:MBO1}
\end{eqnarray}
\begin{eqnarray}
\left.\frac{dV_{\mathrm eff}}{dr}\right|_{r=r_{\mathrm MBO}}=0
\label{eq:MBO2}
\end{eqnarray}

\begin{figure*}[!htp]
\centering
\includegraphics[width=0.95\textwidth]{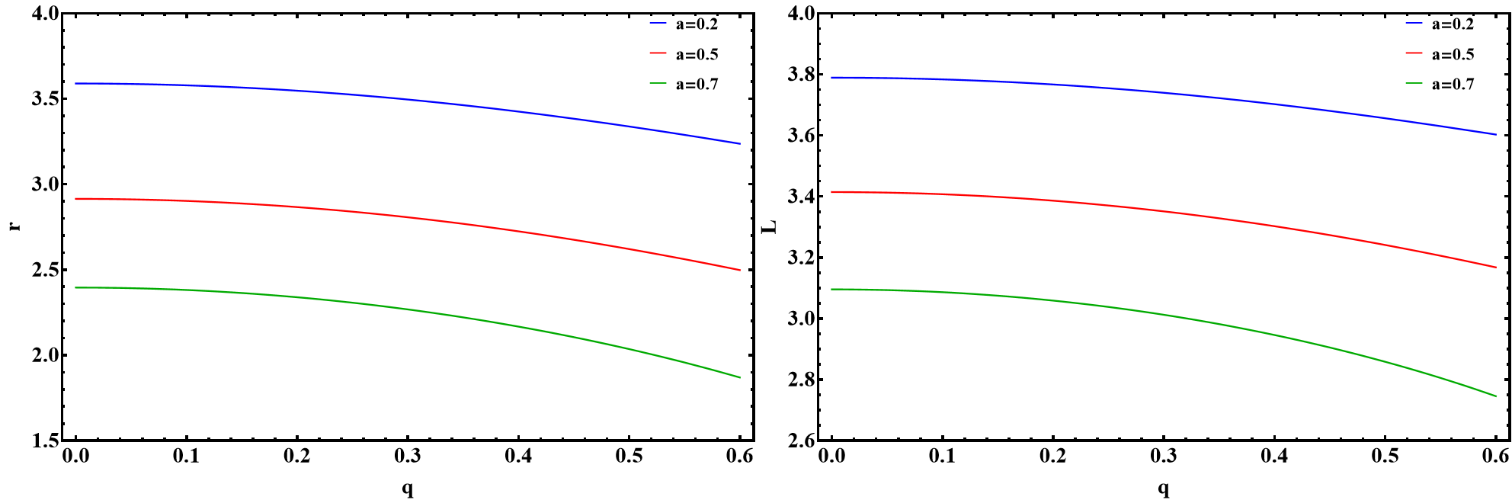}
\caption{Variation of the marginally‑bound‑orbit radius $r_{\mathrm MBO}$ and its angular momentum $L_{\mathrm MBO}$ as the charge parameter changes. Blue: $a = 0.2$; Red: $a = 0.5$; Green: $a = 0.7$.}
\label{fig:fig1}
\end{figure*}

As shown in Fig.~\ref{fig:fig1}, the curves decrease monotonically. With the rotation parameter $a$ held constant, as the charge $q$ increases from $0$ to $0.6$, both $r_{\mathrm MBO}$ and $L_{\mathrm MBO}$ undergo monotonic reduction as charge $q$ grows. For example, when $a = 0.2$, $r_{\mathrm MBO}$ decreases from $3.62$ to $3.26$, and $L_{\mathrm MBO}$ decreases from $3.80$ to $3.60$; when $a = 0.7$, $r_{\mathrm MBO}$ decreases from $2.40$ to $1.87$, and $L_{\mathrm MBO}$ decreases from $3.10$ to $2.75$. Furthermore, a rotation‑enhancement effect can be observed: for a fixed charge parameter $q$, the larger the rotation parameter $a$, the smaller the $r_{\mathrm MBO}$ and $L_{\mathrm MBO}$ become. When $q = 0.6$, the MBO orbital radius at $a = 0.7$ is only $57\%$ of that at $a = 0.2$, indicating a significant synergistic effect between rotation and charge, which together produce inward compression of MBO‑type trajectories toward the black‑hole event horizon. Within the physical range of $q \in [0, 0.6]$, $r_{\mathrm MBO}$ and $L_{\mathrm MBO}$ exhibit an approximately linear relationship with $q$, and the slope increases as the spin parameter $a$ increases. Numerical results indicate that strongly spinning black holes are more sensitive to charge corrections; this type of parameter‑dependent behavior has also been reported in other models of gravitationally corrected black holes (see, e.g., Refs. \cite{45,46}).

The innermost stable circular orbit (ISCO) is the stable inner boundary of a circular orbit; all circular orbits inside this boundary become unstable and rapidly fall into the event horizon. The effective potential simultaneously satisfies the triple stationary point conditions for the function value, first derivative, and second derivative \cite{20}:
\begin{equation}
\begin{split}
V_{\mathrm eff}(r_{\mathrm ISCO}) &= E^2, \\
\left.\frac{dV_{\mathrm eff}}{dr}\right|_{r=r_{\mathrm ISCO}} &= 0, \\
\left.\frac{d^2V_{\mathrm eff}}{dr^2}\right|_{r=r_{\mathrm ISCO}} &= 0.
\end{split}
\label{eq:isco_cond}
\end{equation}

\begin{figure*}[!htp]
\centering
\includegraphics[width=0.98\textwidth]{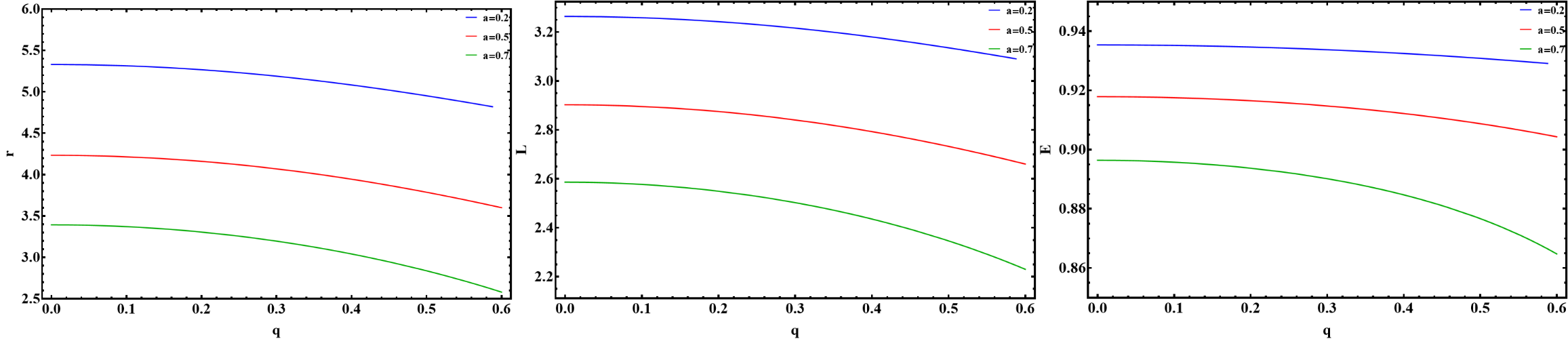}
\caption{How the radius $r_{\mathrm ISCO}$, angular momentum $L_{\mathrm ISCO}$, and energy $E_{\mathrm ISCO}$ of the innermost stable circular orbit vary with the charge parameter. Blue: $a = 0.2$; Red: $a = 0.5$; Green: $a = 0.7$.}
\label{fig:fig2}
\end{figure*}

The ISCO concept and its initial analysis in Kerr black holes were first proposed by Bardeen et al. \cite{20}; Fig.~\ref{fig:fig2} demonstrates how the innermost‑stable‑circular‑orbit radius varies with the charge parameter, with extensions to charged spacetimes (see, e.g., Refs. \cite{44,45}). Numerical results show that, as in the case of MBO orbits, $r_{\mathrm ISCO}$ and $L_{\mathrm ISCO}$ decrease monotonically with increasing $a$ and $q$. When $a = 0.2$, as $q$ increases from $0$ to $0.6$, $r_{\mathrm ISCO}$ decreases from $5.33$ to $4.82$; when $a = 0.7$, $r_{\mathrm ISCO}$ decreases from $3.40$ to $2.52$, showing a more significant contraction; the ISCO energy, $E_{\mathrm ISCO}$, increases monotonically with increasing $a$ and $q$. When $a = 0.2$, $E_{\mathrm ISCO}$ increases from $0.935$ to $0.946$; when $a = 0.7$, $E_{\mathrm ISCO}$ increases from $0.897$ to $0.966$. The increase in energy implies a decrease in the orbital binding energy, making particles more susceptible to perturbations and capture by the black hole. When $q = 0$, the numerical results in this paper are in complete agreement with the classical ISCO results for Kerr black holes ($r_{\mathrm ISCO}= 5.33$ when $a = 0.2$) \cite{20}, verifying the reliability of the numerical calculations. Further numerical studies on the ISCO of charged black holes have been conducted (see, e.g., Refs. \cite{41,46}).

The energy‑angular‑momentum space is an important tool for describing the range of bound orbits (see, e.g., Ref. \cite{25}). In the $(E, L)$ plane, bound orbits correspond to the region bounded by the circular orbit condition $\frac{dV_{\mathrm eff}}{dr}=0$ between the ISCO point ($E = E_{\mathrm ISCO}$, $L = L_{\mathrm ISCO}$) and the MBO point ($E = 1$, $L = L_{\mathrm MBO}$).

\begin{figure*}[!htp]
\centering
\includegraphics[width=0.98\textwidth]{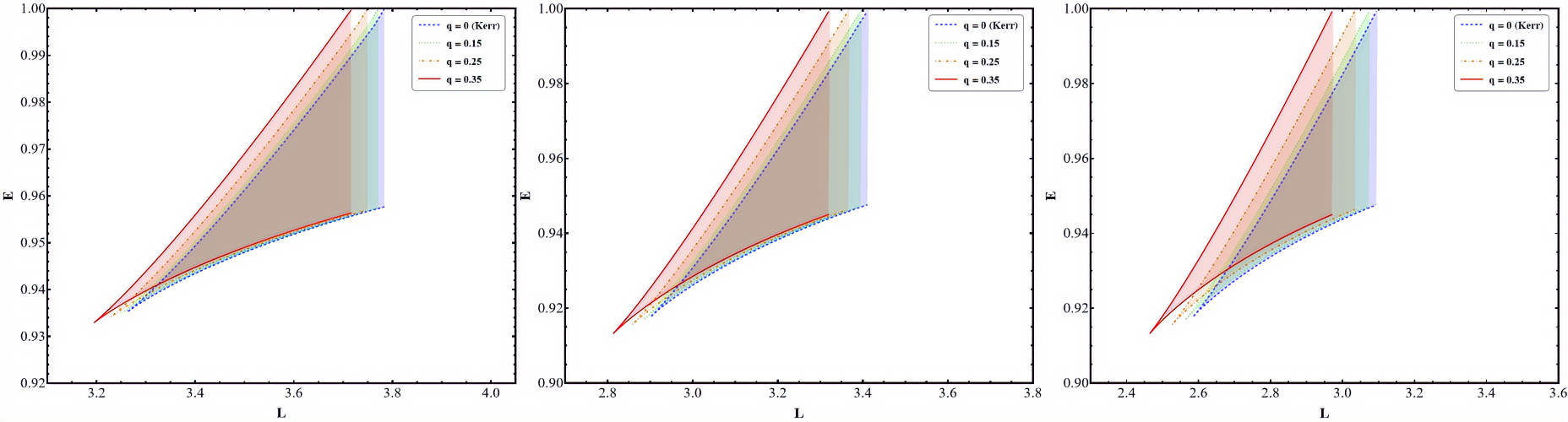}
\caption{The permitted $E$–$L$ regions for particle orbits under different charge parameters $q$ and rotation parameters $a$.}
\label{fig:fig3}
\end{figure*}

Fig.~\ref{fig:fig3} presents a systematic analysis of how different charge and rotation parameters modulate the confined orbits. In this paper, we constructed the allowed regions in $E$‑$L$ phase space for fixed rotation parameters $a = 0.2$, $0.5$, and $0.7$, and charge parameters $q = 0$, $0.15$, $0.25$, and $0.35$. The closed band‑like regions in phase space represent the physically permissible ranges of parameters for bound circular orbits. The boundaries of these regions—comprising the ISCO bifurcation orbit, the branch corresponding to unstable circular orbits, alongside the stable‑circular‑orbit manifold, and the escape critical MBO boundary—together form a regular triangular topological structure.

Phase‑space evolutionary trends illustrate that the charge parameter produces a pronounced squeezing action upon the accessible orbital‑parameter domain. As the NED charge $q$ increases, the overall region of allowed orbits continuously contracts to the left, and the system’s minimum angular momentum ($L_{\mathrm ISCO}$) and maximum bound angular momentum ($L_{\mathrm MBO}$) decrease simultaneously. This indicates that the charge correction weakens the black hole’s gravitational binding range, allowing particles to form bound orbits only within a narrower angular momentum range. Second, the black hole’s spin amplifies the modulating effect of charge. Under high spin conditions ($a = 0.7$), the phase space contraction is most pronounced, and the allowed orbital range narrows significantly; whereas in low‑spin scenarios ($a = 0.2$), the phase space structure more closely resembles that of a classical Kerr black hole. The coupling between spin and charge renders the orbital parameter space of NED black holes highly tunable. The Kerr limit with zero charge ($q = 0$) always serves as the reference structure, featuring the largest phase space and the richest set of bound orbits. With the introduction of charge, the boundaries of the phase space undergo a systematic shift, demonstrating that nonlinear electrodynamic corrections fundamentally alter the conditions for the existence of particle orbits in strong gravitational fields (see, e.g., Refs. \cite{47,59}).

Overall, the boundary conditions in the $E$‑$L$ phase spaces provide strict physical constraints for the subsequent selection of parameters for $(z, w, v)$ resonant periodic orbits; relevant analysis of the Kerr black hole phase space (see, e.g., Refs. \cite{37,54}).

Once the effective potential has been constructed, one may explore accessible test‑particle trajectories within black‑hole gravity by means of the radial equation of motion $\dot{r}^2=E^2-V_{\mathrm eff}(r)$. For a given energy $E$ and angular momentum $L$, orbits can be classified into four types—absorption orbits, bound orbits, MBO orbits, and scattering orbits—based on the number of real roots of $\dot{r}^2=0$ (see, e.g., Ref. \cite{43}). Fixing $a = 0.5$ and $q = 0.25$, we generate plots showing $\dot{r}^2$ against radial coordinate $r$ for a set of distinct particle‑energy values $E$. Such numerical investigations offer key insight into the full dynamical pathways followed by test bodies: these objects may plunge inward starting from spatial infinity, get gravitationally captured by the black‑hole potential well, or sustain periodic motion along bound trajectories.

Orbital types are classified as follows:

Absorption orbit: When $E < E_{\mathrm ISCO} = 0.917107$, $\dot{r}^2=0$ has only one positive real root. The test‑particle plunges straight through the event‑horizon region, with no possibility of establishing gravitationally bound trajectories.

Bound orbits: When $E_{\mathrm ISCO} < E < 1$, whenever the radial‑discriminant equation yields two distinct positive real roots, these solutions correspond to the inner turning point (periastron) and outer turning point (apastron). Test‑particle trajectories oscillate between these two radial boundaries and thus realise gravitationally bound orbits. For example, when $E = 0.965$, the particle moves between $r_1 \approx 2.1$ and $r_2 \approx 8.5$.

Marginally‑bound orbit: When $E = 1$, the two radial turning points coincide, and the radial period tends to infinity, corresponding to an MBO orbit.

Scattering orbit: When $E > 1$, $\dot{r}^2=0$ has only one positive real root, and the particle will escape to infinity after grazing the black hole.

\begin{figure*}[!htp]
\centering
\includegraphics[width=0.95\textwidth]{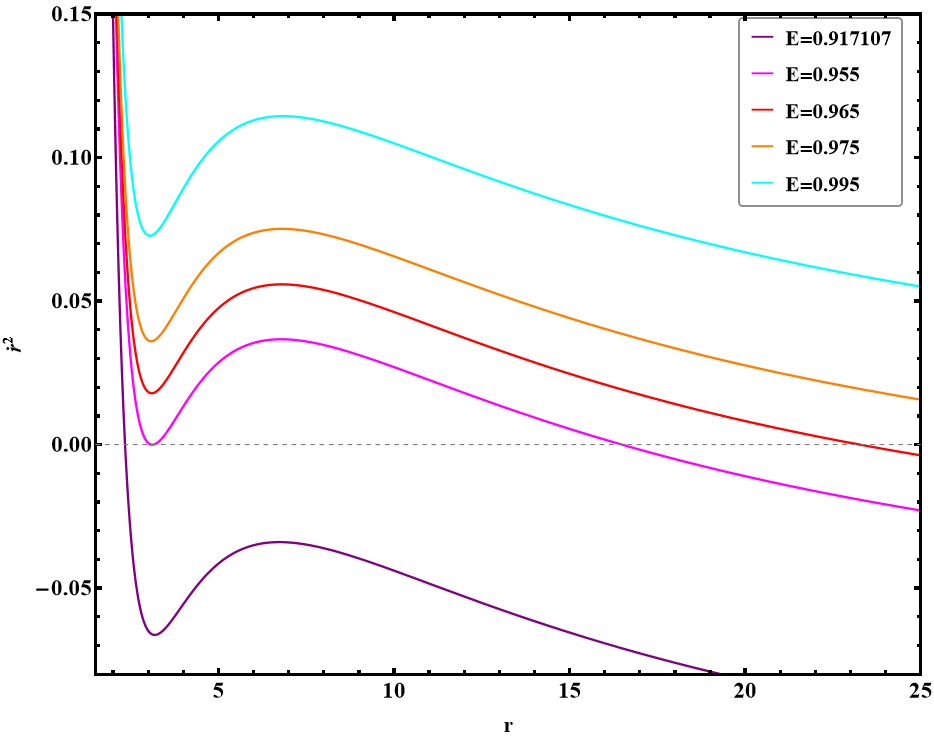}
\caption{Equation of motion $\dot{r}^2$ for particles around a rotating, charged, nonlinear electrodynamic black hole as a function of the radial coordinate $r$ ($a = 0.5$, $q = 0.25$). Purple: $E = 0.917107$ (ISCO); Magenta: $E = 0.955$; Red: $E = 0.965$; Orange: $E = 0.975$; Cyan: $E = 0.995$.}
\label{fig:fig4}
\end{figure*}

To further elucidate the details of single‑particle orbital dynamics, we fixed typical spacetime parameters $a = 0.5$, $q = 0.25$, and particle energy $E = 0.96$, and compared the effective potential distributions corresponding to different orbital angular momenta $L$. The results show that the effective potential of a NED black hole still retains a classical bipolar structure with a single maximum and a single minimum, which separate unstable from stable circular orbital families, respectively. Growing values of orbital‑angular‑momentum $L$ produce a global upward shift for the effective‑potential curve, the height of the potential barrier increases significantly, and the positions of the extrema shift radially outward. From a physical standpoint, elevated orbital‑angular‑momentum values amplify the centrifugal‑potential contribution governing test‑particle motion, which shifts the trajectory outward to larger radial separations from the black‑hole event horizon. Under low‑angular‑momentum conditions, the potential well is closer to the event horizon, and strong gravitational corrections become more pronounced.

A gravitationally bound trajectory can form provided that the test‑particle energy sits between the local minimum and local maximum values of the effective potential. At this point, the radial discriminant function has two zeros, and the particle undergoes radial oscillations back and forth between the inner and outer turning points. When the energy equals the extreme energy values, the radial velocity is zero, and the orbit degenerates into a circular orbit; the point where the extreme values coincide corresponds to the critical ISCO orbit, which serves as the inner boundary of stable bound orbits. Compared to Kerr spacetime, the NED charge correction lowers the effective potential barrier, making it easier for particles to form narrow‑amplitude oscillating orbits near the black hole and providing the dynamical basis for the formation of higher‑order resonant periodic orbits (see, e.g., Ref. \cite{39}).

\begin{figure*}[!htp]
\centering
\includegraphics[width=0.95\textwidth]{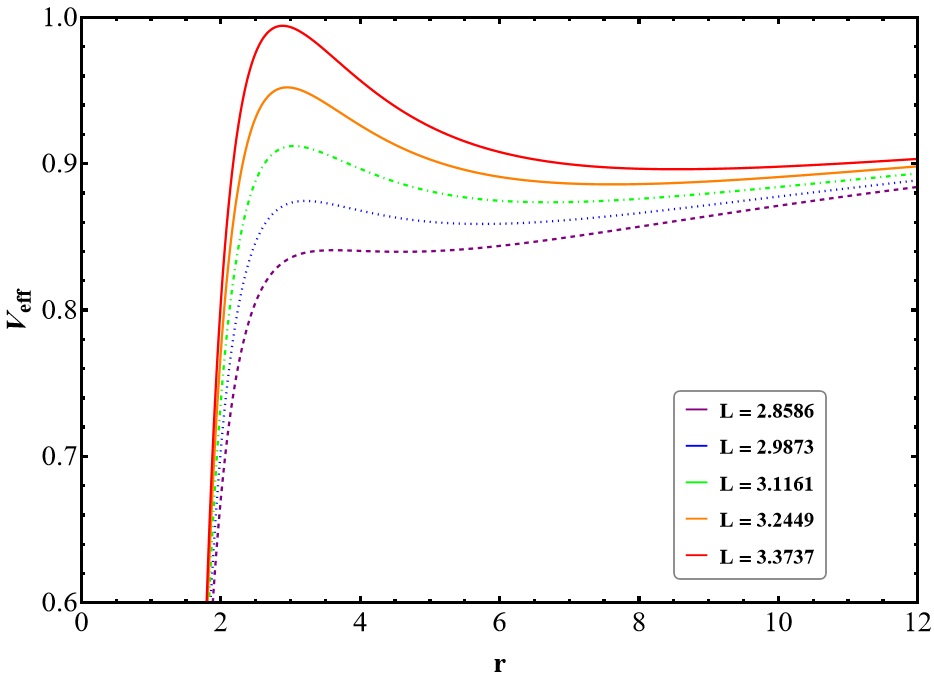}
\caption{How the effective potential varies with $r$ for different angular momenta. Here, the parameters $a = 0.5$ and $q = 0.25$ are fixed.}
\label{fig:fig5}
\end{figure*} 

\section{Periodic Orbital Dynamics}\label{sec4}

Within backgrounds described by rotating‑charged NED black‑hole geometries, one may sort bound test‑particle trajectories using the ratio between azimuthal and radial oscillation frequencies. Irrational values for this ratio produce orbits displaying ongoing precessional motion; rational ratios yield trajectories that close upon themselves to generate periodic‑orbit solutions (see, e.g., Ref. \cite{55}). In this section, we use a rational number coding method to classify periodic orbits, numerically construct periodic orbits under different parameter combinations, explore how spin parameter $a$ and charge parameter $q$ shape the orbital‑energy properties, angular momentum, and geometric shape, and investigate the evolution of zoom‑whirl behavior as a function of these parameters.

During a single complete radial round trip ($r_1\to r_2\to r_1$) of a particle, the total azimuthal displacement is given by the radial integral:
\begin{equation}
\begin{split}
\Delta\phi &= 2\int_{r_1}^{r_2}\frac{d\phi}{dr}\,dr \\
&= 2\int_{r_1}^{r_2}\frac{L-aE}{r^2\sqrt{E^2-V_{\mathrm eff}(r)}}\,dr
\end{split}
\label{eq:23}
\end{equation}
Here, $\Delta\phi$ is the angular increment swept by the particle during one complete radial cycle. Quantities $r_1$ and $r_2$ correspond to the two radial‑motion turning points, which are solved from the two real roots of the condition $\dot{r}^2=0$.

The radial‑azimuthal precession ratio $Q$ (the number of azimuthal rotations within a single radial period) is defined as:
\begin{align}
Q=\frac{\Delta\phi}{2\pi}-1=w+\frac{v}{z}
\label{eq:Qdef}
\end{align}

For irrational values of the resonance parameter $Q$, neutral test‑particle trajectories undergo precessional motion, an effect analogous to Mercury’s perihelion shift within our Solar System. When $Q$ is a rational number, the orbit exhibits periodic closed characteristics and can be uniquely identified by three integers $(z, w, v)$, in which integer $z$ quantifies the total count of orbital lobes, while $w$ denotes the zoom‑phase index, and $v$ the whirl number, following the $(z,w,v)$ taxonomy of Levin \& Perez‑Giz \cite{34}. Together these three integers capture geometric traits for test‑particle trajectories evolving within strong‑gravity backgrounds. Specifically, the zoom phase describes the comparatively sluggish particle motion taking place in the vicinity of the apocenter, and the whirl phase corresponds to the high‑curvature trajectory as the particle rapidly sweeps past the pericenter. Various extensions of this coding system are discussed subsequently (see, e.g., Refs. \cite{35,36}).

In this section, we select three sets of spin parameters, $a = \{0.2, 0.5, 0.7\}$, and four sets of NED charges, $q = \{0, 0.15, 0.25, 0.35\}$, to plot surfaces showing how the precession ratio $Q$ varies with energy $E$ and angular momentum $L$, thereby quantitatively revealing the synergistic modulation effect of spin and NED charge on orbital resonance characteristics.

The top subfigure shows the evolution of resonance‑precession ratio $Q$ with orbital energy $E$ per unit mass. The horizontal axis represents the particle’s conserved energy $E$, and the vertical axis represents the precession ratio $Q$. Within the confinement range $E_{\mathrm ISCO} < E < 1$, where $a$ and $q$ are fixed, $Q$ rises monotonically and rapidly as $E$ increases; as $E$ approaches $1$ near the MBO boundary, the curve diverges steeply. Physical interpretation: As the particle’s energy approaches the escape threshold, the apocenter $r_2$ extends outward, and the number of precession cycles per radial round trip increases significantly. For a fixed spin $a$ and the same energy $E$, the larger the charge $q$, the smaller the precession ratio $Q$; the NED correction term $\frac{q^4}{r^4}$ compresses the radial activity interval $[r_1,r_2]$, suppressing the rate of azimuthal precession. Thus, at the same energy, the higher the charge, the smaller the azimuthal rotation per radial oscillation. Comparing the three sets of subplots, as spin $a$ increases, the vertical spacing between the curves for different charges widens, and the modulatory effect of NED charge on the precession ratio is amplified by spin. At low spin ($a = 0.2$), the four curves lie closely together, while at high spin ($a = 0.7$), the degree of stratification is significantly enhanced.

The lower subfigure illustrates how the precession ratio $Q$ varies with the specific orbital‑angular‑momentum quantity $L$. The horizontal axis represents the conserved orbital‑angular‑momentum quantity $L$, and the vertical axis represents the precession ratio $Q$. When $a$ and $q$ are fixed, $Q$ monotonically decays as $L$ increases, and $Q$ rises sharply as $L$ approaches the low‑angular‑momentum region near ISCO ($L\to L_{\mathrm ISCO}$). Physical interpretation: Trajectories carrying small orbital‑angular‑momentum values reside in near‑horizon regions, where pronounced spacetime distortions amplify the rate of azimuthal precessional motion; high‑angular‑momentum orbits are far from the event horizon, where spacetime becomes nearly flat, and the precession effect weakens. With spin $a$ fixed, as the NED charge $q$ increases, the entire curve shifts toward lower angular momenta, while simultaneously compressing downward. Taking $a = 0.2$ as an example: the curve for $q = 0$ lies in the interval $L \in [3.45, 3.75]$, while the curve for $q = 0.35$ contracts to a smaller angular momentum range, indicating that charge compresses the range of angular momentum values allowed by the bound orbit. As spin $a$ increases, all curves shift significantly to the left, and the valid angular momentum range narrows overall; at $a = 0.7$, the lower limit of the minimum angular momentum drops to $L \approx 2.90$, which is much smaller than the $L \approx 3.45$ observed at $a = 0.2$.

\begin{figure*}[!htp]
\centering
\begin{minipage}{\textwidth}
  \centering
  \includegraphics[width=0.95\textwidth]{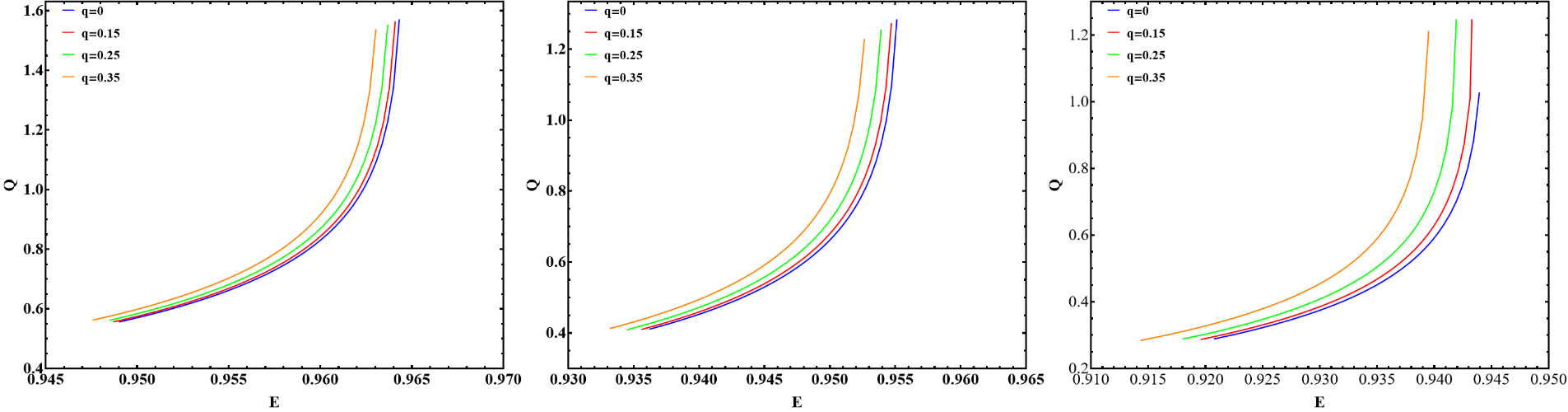}
  \subcaption*{(a) Precession ratio $Q$ as function of orbital energy $E$, with $L$ fixed at $\frac{L_{\mathrm ISCO}+L_{\mathrm MBO}}{2}$.}
\end{minipage}

\vspace{8pt} 

\begin{minipage}{\textwidth}
  \centering
  \includegraphics[width=0.95\textwidth]{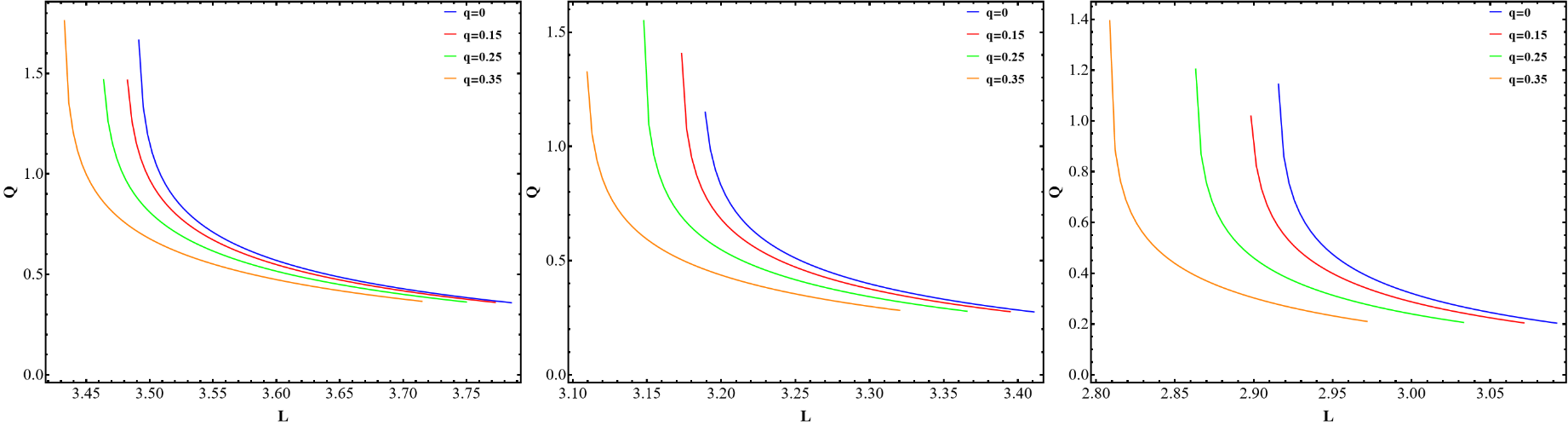}
  \subcaption*{(b) Precession ratio $Q$ versus angular momentum $L$, for fixed $E=0.96$.}
\end{minipage}

\caption{$Q$ as a function of $E$ or $L$. Panel (a): $Q$‑$E$ relation; Panel (b): $Q$‑$L$ relation. Spin parameters $a=\{0.2,0.5,0.7\}$, charge values $q=\{0,0.15,0.25,0.35\}$.}
\label{fig:fig6}
\end{figure*}

A resonant periodic orbit $(z, w, v)$ satisfies $Q = w+\frac{v}{z}$. During a complete resonance period, the particle completes $z$ radial round trips, and the closed orbit forms $z$ outer symmetric lobes, where integer index $z$ gives the lobe count, which exactly matches the total number of radial oscillation cycles completed over one full orbital period. With the spin fixed at $a = 0.2$, compare the orbital plane geometry of the charge‑free Kerr case ($q = 0$, blue solid line) with that of the NED charge case ($q = 0.25$, red dashed line).

\begin{table*}[!htp]
\centering
\setlength{\tabcolsep}{3pt} 
\footnotesize
\caption{When $a = 0.2$, $(z, w, v)$ represent the numerical values of the energy $E$ for different periodic orbits. The charge parameter $q$ varies from $0$ to $0.35$. Here, $L_{av}=\frac{L_{\mathrm ISCO}+L_{\mathrm MBO}}{2}$.}
\label{tab:tab1}
\begin{tabular*}{\textwidth}{@{\extracolsep\fill}lccccccccccccc}
\toprule
$q$ & $L_{av}$ & $E_{(1,1,0)}$ & $E_{(1,2,0)}$ & $E_{(1,3,0)}$ & $E_{(2,1,1)}$ & $E_{(2,2,1)}$ & $E_{(2,3,1)}$ & $E_{(3,1,2)}$ & $E_{(3,2,2)}$ & $E_{(3,3,2)}$ & $E_{(4,0,3)}$ & $E_{(4,1,3)}$ & $E_{(4,2,3)}$ \\
\midrule
0     & 3.5266 & 0.9524 & 0.9584 & 0.9652 & 0.9613 & 0.9535 & 0.9581 & 0.9511 & 0.9612 & 0.9551 & 0.9611 & 0.9518 & 0.9612 \\
0.15  & 3.5140 & 0.9769 & 0.9580 & 0.9653 & 0.9626 & 0.9657 & 0.9582 & 0.9554 & 0.9624 & 0.9553 & 0.9541 & 0.9657 & 0.9624 \\
0.25  & 3.4916 & 0.9783 & 0.9583 & 0.9654 & 0.9818 & 0.9766 & 0.9584 & 0.9730 & 0.9650 & 0.9818 & 0.9628 & 0.9783 & 0.9650 \\
0.35  & 3.4572 & 0.9830 & 0.9594 & 0.9656 & 0.9690 & 0.9940 & 0.9587 & 0.9744 & 0.9690 & 0.9758 & 0.9642 & 0.9986 & 0.9690 \\
\botrule
\end{tabular*}
\end{table*}

\begin{table*}[!htp]
\centering
\setlength{\tabcolsep}{3pt}
\footnotesize
\caption{When $a = 0.2$, $(z, w, v)$ represent the values of angular momentum $L$ for different periodic orbits. The charge parameter $q$ varies from $0$ to $0.35$. Here, $E = 0.96$.}
\label{tab:tab2}
\begin{tabular*}{\textwidth}{@{\extracolsep\fill}lccccccccccccc}
\toprule
$q$ & $L_{(1,1,0)}$ & $L_{(1,2,0)}$ & $L_{(1,3,0)}$ & $L_{(2,1,1)}$ & $L_{(2,2,1)}$ & $L_{(2,3,1)}$ & $L_{(3,1,2)}$ & $L_{(3,2,2)}$ & $L_{(3,3,2)}$ & $L_{(4,0,3)}$ & $L_{(4,1,3)}$ & $L_{(4,2,3)}$ \\
\midrule
0     & 3.6169 & 3.8100 & 4.1000 & 3.5010 & 3.6500 & 3.8000 & 3.5816 & 3.5000 & 3.6999 & 3.4991 & 3.6000 & 3.5000 \\
0.15  & 3.6106 & 3.7938 & 4.1000 & 3.5002 & 3.5250 & 3.8000 & 3.4424 & 3.4988 & 3.7000 & 3.4314 & 3.5250 & 3.4988 \\
0.25  & 3.6000 & 3.7974 & 4.1000 & 3.6250 & 3.5875 & 3.8000 & 3.5608 & 3.4999 & 3.6250 & 3.4828 & 3.6000 & 3.4999 \\
0.35  & 3.6010 & 3.8250 & 4.1000 & 3.5001 & 3.6499 & 3.8000 & 3.5394 & 3.4997 & 3.5500 & 3.4641 & 3.6000 & 3.4997 \\
\botrule
\end{tabular*}
\end{table*}

The NED correction term $\frac{2 q^4}{5 r^4}$ significantly alters the spacetime curvature only in the region near the event horizon where $r$ is small; at large distances where $r \to \infty$, the correction term rapidly decays to $0$. Therefore, the modulation of the orbit by the charge is local, and the effective potential is raised overall. At a given $r$, an increase in $q$ raises $V_{\mathrm eff}(r)$, while the upper bound of the bound orbit energy remains constant ($E < 1$). Resonance matching can only be achieved by reducing the angular momentum $L$. The width of the effective potential well narrows as $q$ increases; $r_1$ increases slightly while $r_2$ decreases significantly, narrowing the range of radial oscillations and reducing the total precession per round trip, causing the precession ratio $Q$ to decrease accordingly. Elevated black‑hole spin $a$ raises mean curvature along test‑particle trajectories and magnifies curvature corrections induced by NED physics in the near‑horizon domain; at high spin, the modulation of the charge upon precession ratio $Q$ and orbital geometry is far greater than at low spin ($a = 0.2$). Similar spin‑parameter coupling effects have also been observed in other gravitationally modified black holes (see, e.g., Refs. \cite{47,55}).

\begin{figure*}[!htp]
\centering

\begin{minipage}{\textwidth}
  \centering
  \includegraphics[width=0.95\textwidth]{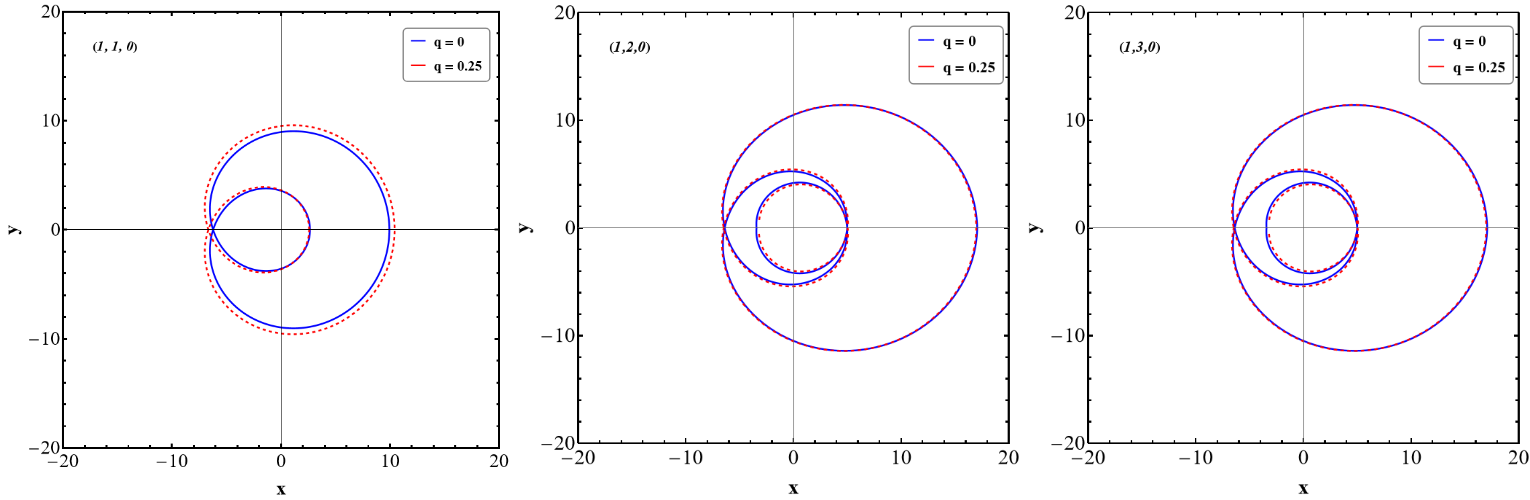}
\end{minipage}
\vspace{8pt}

\begin{minipage}{\textwidth}
  \centering
  \includegraphics[width=0.95\textwidth]{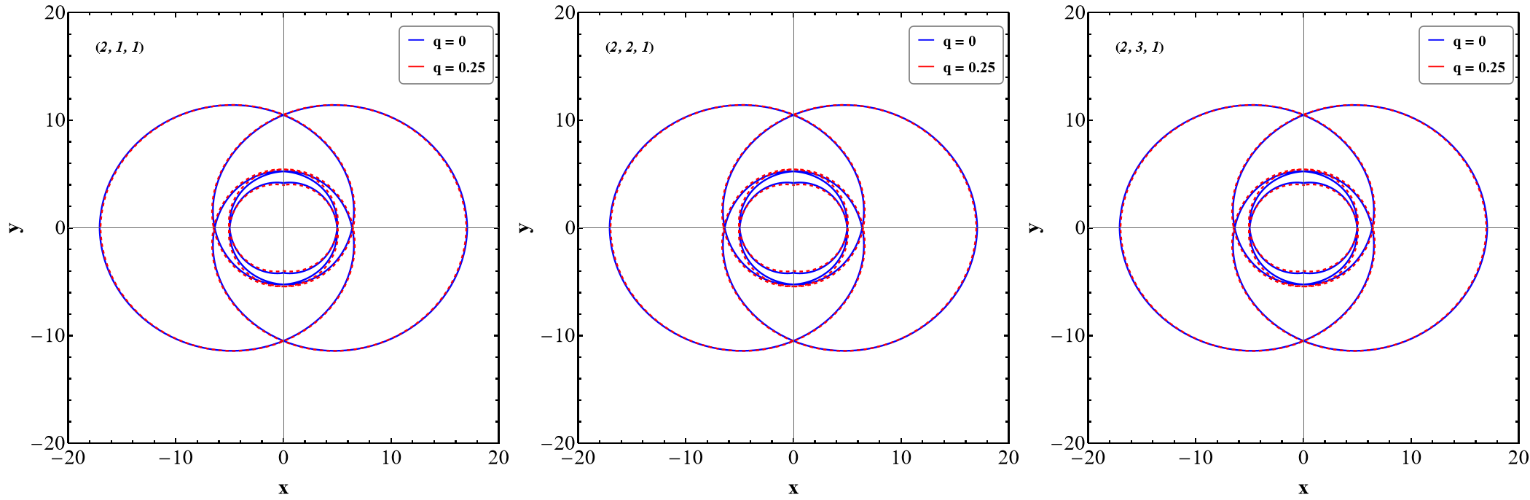}
\end{minipage}
\vspace{8pt}

\begin{minipage}{\textwidth}
  \centering
  \includegraphics[width=0.95\textwidth]{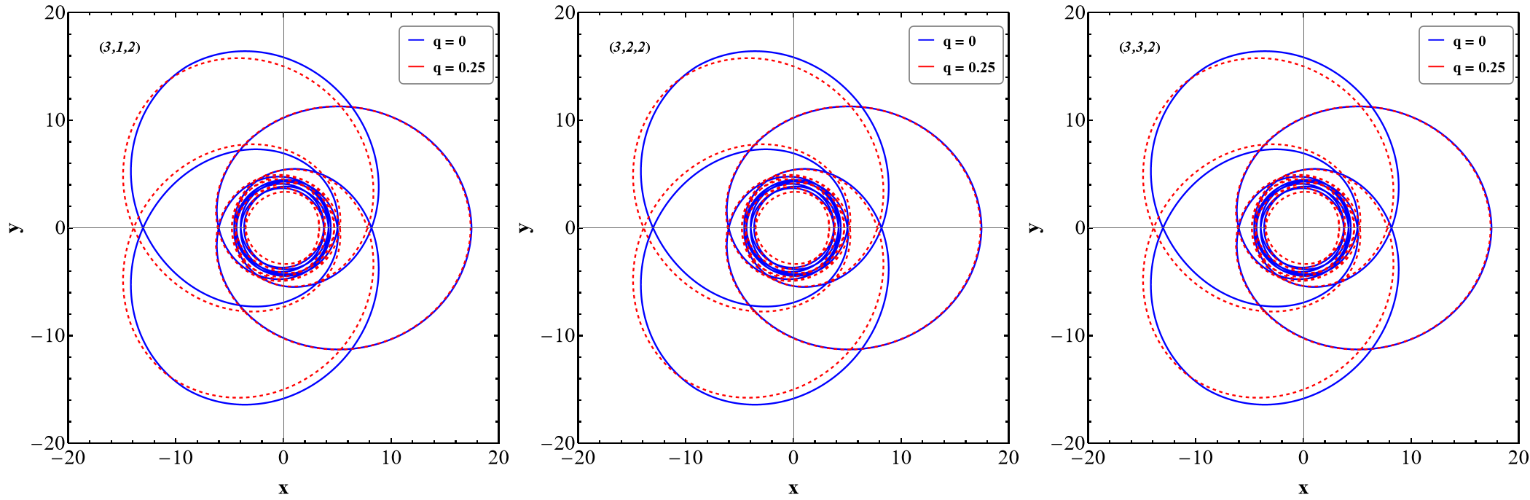}
\end{minipage}
\vspace{8pt}

\begin{minipage}{\textwidth}
  \centering
  \includegraphics[width=0.95\textwidth]{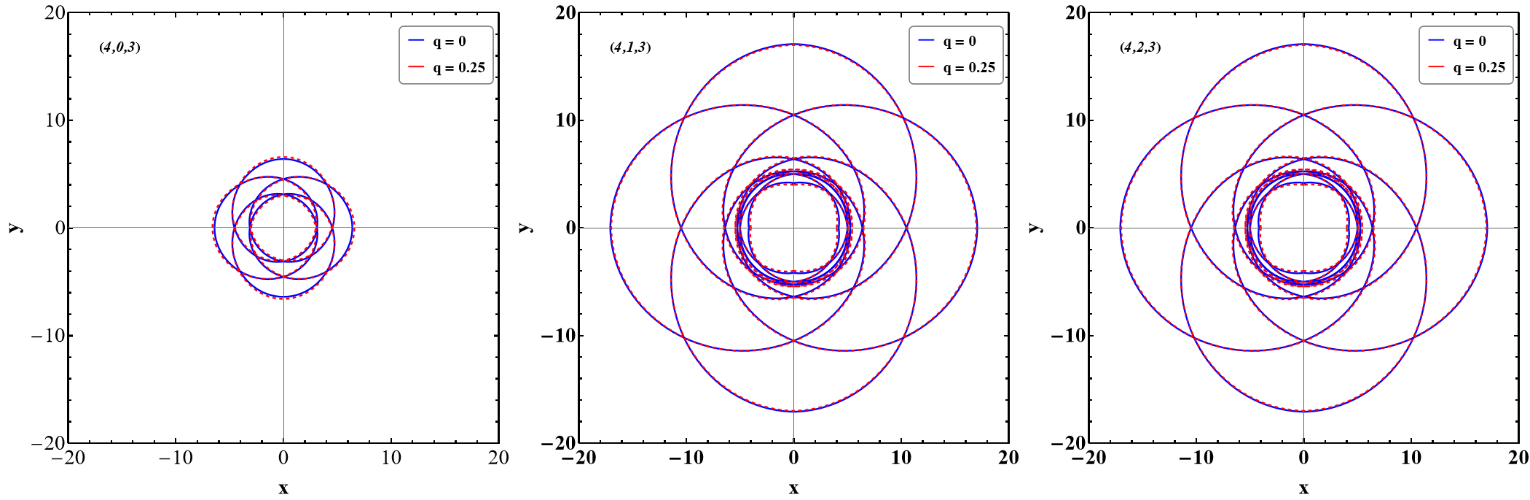}
\end{minipage}

\caption{Trajectories of periodic orbits $(z,w,v)$ for a neutral test body in rotating charged NED black hole spacetime. Here $a = 0.2$, $q = 0,0.25$, $E = 0.96$; angular momentum values are given in Table~\ref{tab:tab2}. The coordinate transformation reads $(x,y)=\left(r\cos\phi, r\sin\phi\right)$.}
\label{fig:fig7}
\end{figure*}

\section{Gravitational Wave Radiation from Periodic Orbits}\label{sec5}

In the extreme‑mass‑ratio inspiral (EMRI) systems studied in this paper, a rotating, charged NED black hole serves as the massive primary black hole, with a compact secondary body serving as the test‑particle component. This compact object follows periodic bound trajectories through the background’s strong‑gravity domain and radiates gravitational‑wave energy. Given the substantial mass disparity between this orbiting compact body and the central NED black hole, the small compact object contributes negligible perturbation to the background spacetime, so we adopt a fixed‑background geodesic approximation. The evolution of spacetime is entirely governed by the metric of a static charged NED black hole. Therefore, energy and angular momentum losses for the system remain small across one or multiple orbital cycles. This property allows orbital dynamical evolution to follow the adiabatic approximation. Within this approximation scheme, a particle’s conserved energy and orbital angular momentum may be treated as constant over a full orbital cycle. Gravitational‑wave radiation waveforms can then be derived from the periodic motion features of geodesic orbits. This approach is the standard method for modeling gravitational waves in systems with extreme‑mass‑ratios, and it has been extensively validated and applied in numerous related studies (see, e.g., Refs. \cite{35,41,58}).

To precisely compute gravitational‑wave signals generated by periodic orbits around charged black‑hole backgrounds, this paper employs the Kludge waveform method for waveform construction. This model was proposed by Babak et al. \cite{38}, and its early prototypes and subsequent improvements (see, e.g., Refs. \cite{22,42}) are based on EMRI gravitational‑wave modeling using a general parameterized metric (see, e.g., Ref. \cite{36}). This method first relies on the Kerr‑NED black hole geodesic equations derived earlier to solve for the complete periodic orbital evolution of a particle, obtaining information on the particle’s radial position and azimuthal phase at any given time. It then performs a second‑order approximation based on the general relativity quadrupole radiation formula, ultimately decomposing the signal into a bipolarized gravitational wave signal that can be directly detected by a detector. The quadrupole radiation formalism was originally derived by Peters \cite{39}; implementations tailored for EMRI systems can be found (see, e.g., Refs. \cite{22,57}).
\begin{align}
h_{ij}= \frac{4\eta M}{D_L}\left(v_i v_j-\frac{m}{r} n_i n_j\right)
\label{eq:quad}
\end{align}
In this expression, $M$ denotes the mass of the central rotating electrically‑charged NED black hole, and $m$ denotes the mass of the orbiting compact object that satisfies the extreme mass ratio condition $m\ll M$. Here $D_L$ stands for the luminosity distance to the source system, $\eta=\frac{Mm}{(M+m)^2}$ is the symmetric mass ratio for the extreme‑mass‑ratio configuration, $v_i$ gives velocity vector components for the test‑particle orbital motion and describes instantaneous orbital states, while $n_i$ represents the unit radial vector pointing outward from the central black hole toward the compact object. The quadrupole‑moment relation completely captures spacetime metric perturbations sourced by periodic orbital motion within the strong gravity of a charged black hole, and it illustrates core distribution properties of the emitted gravitational‑wave radiation (see, e.g., Refs. \cite{51,52}).

To align the observational coordinate systems of ground‑based and space‑based gravitational wave detectors, the raw spacetime perturbation signals obtained from the quadrupole moment solution must be projected onto the detector’s orthogonal observation basis vectors to perform a dual‑polarization component decomposition. We adopt a standard detector‑adapted coordinate system, define the periapsis longitude $\omega$ and the orbital inclination angle $\iota$, and construct three sets of orthogonal observational basis vectors (see, e.g., Refs. \cite{41,58}).
\begin{align}
e_X&=\left[\cos\omega,\,-\sin\omega,\,0\right]
\label{eq:ex}\\
e_Y&=\left[\cos\iota\sin\omega,\,\cos\iota\cos\omega,\,-\sin\iota\right]
\label{eq:ey}\\
e_Z&=\left[\sin\iota\sin\omega,\,\sin\iota\cos\omega,\,\cos\iota\right]
\label{eq:ez}
\end{align}

The aforementioned orthogonal basis vectors form the standard projected coordinate system for gravitational wave observations, enabling the precise decomposition of the spacetime perturbation tensor into two orthogonal and independent polarization modes: the ``$+$'' and ``$\times$'' polarizations of gravitational waves. Based on this observational coordinate system and the theory of second‑order quadrupole radiation, two sets of analytical expressions for gravitational wave polarization corresponding to periodic orbits in the Kerr‑NED charged black hole spacetime were ultimately derived (see, e.g., Refs. \cite{57,58}).
\begin{align}
h_+&= -\frac{2\eta M^2}{D_L\,r}\left(1+\cos^2\iota\right)\cos\left(2\phi+2\zeta\right)
\label{eq:hplus}\\
h_\times&= -\frac{4\eta M^2}{D_L\,r}\cos\iota\,\sin\left(2\phi+2\zeta\right)
\label{eq:hcross}
\end{align}
In this equation, $\phi$ represents the instantaneous azimuthal phase of the particle’s orbit, $\zeta$ represents the initial phase shift of the orbit, and the inclination parameter $\iota$ characterizes the inclination angle separating the orbital plane from the observer’s line of sight. As can be seen from the waveform expression, the gravitational‑wave radiation amplitude is inversely proportional to the particle’s instantaneous orbital radius $r$. When the particle reaches the periastron, the radial radius is minimal, and the radiation amplitude significantly increases, forming a pulse peak; as the particle moves toward the apastron, the radiation amplitude decays, ultimately forming the characteristic periodic‑orbital gravitational‑wave waveform—a low‑frequency envelope modulating a high‑frequency pulse. At the same time, $h_+$ and $h_\times$ strictly maintain an orthogonal phase relationship, consistent with the fundamental laws of gravitational wave polarization in general relativity.

This paper fixes observational and fundamental astronomical quantities for extreme mass ratio systems, so as to systematically explore how ``hair‑related'' parameters including black‑hole spin and NED charge shape gravitational‑wave radiative properties of periodic orbits. In particular, we assign $M=10^{7}M_\odot$ for the central rotating charged black‑hole solution described within nonlinear electrodynamics (NED), and specify $m=10M_\odot$ as the mass belonging to the compact orbiting test object, the observation inclination and initial phase are set to $\iota=\frac{\pi}{4}$ and $\zeta=\frac{\pi}{4}$, respectively, and the system’s observational luminosity distance is fixed at $D_L = 200$ Mpc. Based on these fixed observational and celestial body mass parameters, by varying the NED charge parameters and black hole spin parameters, we can quantitatively analyze the regulatory mechanisms by which the ``hair'' features of charged black holes modulate the evolution of gravitational wave amplitude, phase, and envelope, as well as the density of pulses. This enables a comparative distinction between the gravitational wave observational characteristics of the modified gravitational spacetime and those of the classical hairless Kerr spacetime.

With a fixed black hole spin of $a = 0.5$ and a NED charge of $q = 0.25$, we compare the gravitational wave evolution curves for the three types of resonant orbits: $(1,2,0)$, $(2,2,1)$, and $(3,1,2)$. The overall differences in the waveforms are primarily determined by the orbital precession ratio $Q$ and the number of radial periods. For the $(1,2,0)$ orbit, the precession ratio $Q = 2.0$ and the radial period $z = 1$; a single radial round trip corresponds to an azimuthal rotation angle of $4\pi$. This orbit has the fastest precession rate, a higher orbital eccentricity, an extremely small periastron radial scale, densely packed pulse clusters, and the highest peak radiation amplitude. Due to the constraints of the numerical integration truncation rule, the orbital evolution covers only the $0$–$300$-second time window, with no radiation signal observed in the latter half of the observation interval. For the $(2,2,1)$ orbit, the precession ratio $Q = 2.5$, the orbital precession rate is moderate; the density of pulses and the peak amplitude both fall between those of the other two sets of topological orbits; the duration of the complete orbital evolution is slightly longer, with the waveform extending to approximately $350$ seconds; similarly, early truncation of the integration occurs (see, e.g., Refs. \cite{37,48}). For the $(3,1,2)$ orbital precession ratio $Q = \frac{5}{3}\approx 1.667$, it exhibits the slowest orbital precession rate, the longest time span corresponding to a complete resonance period, and its orbital evolution fully spans the entire observation interval from $0$ to $1100$ seconds. The low‑frequency slow envelope exhibits gentle fluctuations, the time intervals between pulse clusters are wider, and complete pulse groups appear at the $400$‑second, $700$‑second, and $1000$‑second marks. In this numerical simulation, all integrations are initialized at the periastron of each orbit; the initial radial coordinate was set to its minimum value, the azimuth was set to zero, and the global phase $\zeta=0$. At the initial time, the radiation amplitudes $h_+$ and $h_\times$ were simultaneously reset to zero, and the three topological orbital curves intersected at the origin on the $\tau$‑axis at $\tau=0$.

\begin{figure*}[!htp]
\centering
\includegraphics[width=0.95\textwidth]{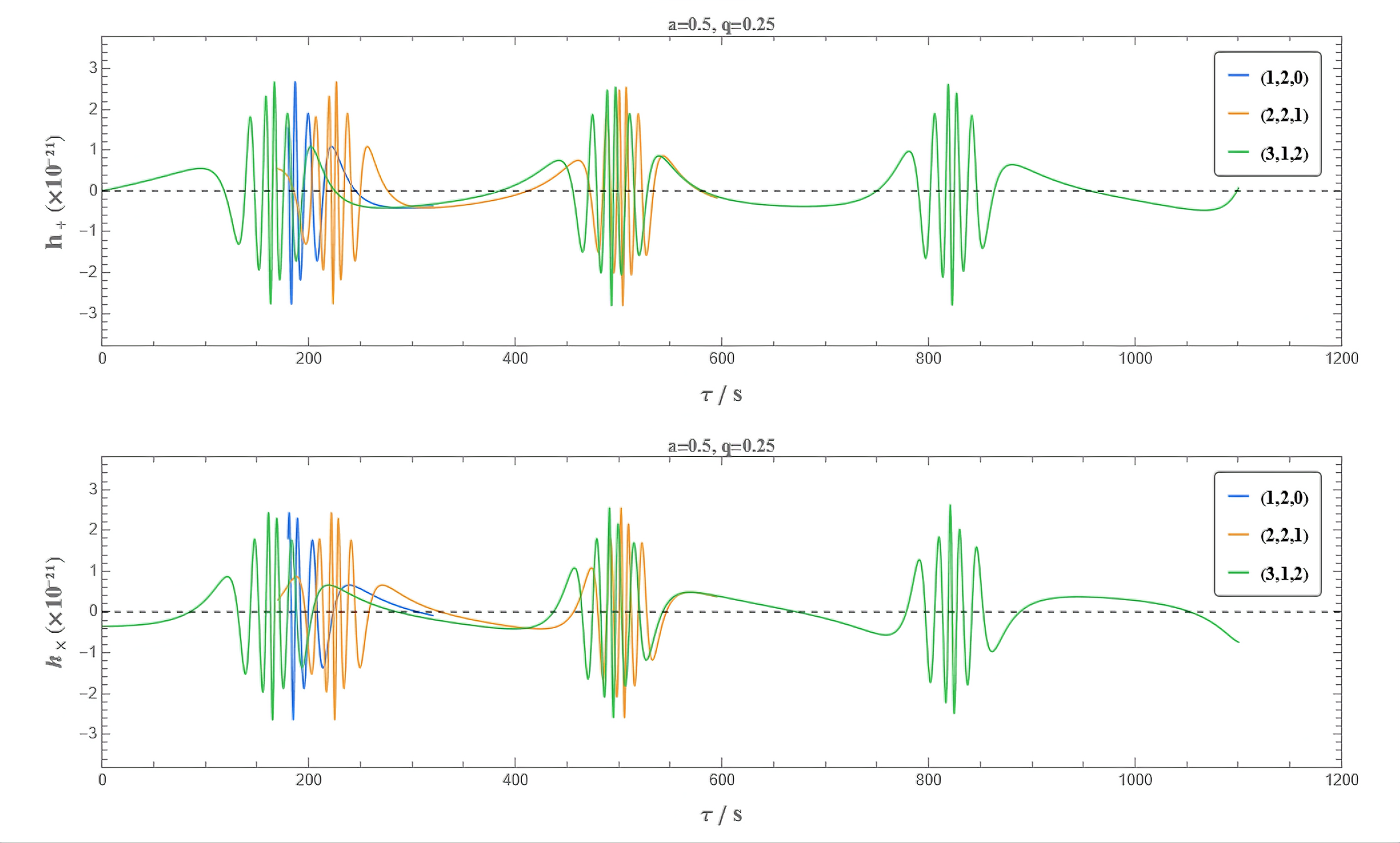}
\caption{Gravitational‑wave waveforms generated by periodic orbits with different topologies. The top and bottom panels correspond to different polarization states. The top figure shows $h_+$, and the bottom figure shows $h_\times$, with the parameters set to $a = 0.5$ and $q = 0.25$.}
\label{fig:fig8}
\end{figure*}

For the fixed resonant topology $(3,1,2)$ and a black hole spin of $a = 0.5$, varying the NED charge parameter $q$ allows the range of orbital radial oscillations to be adjusted via the effective spacetime potential correction term $\frac{2 q^4}{5 r^4}$, thereby modulating the gravitational waveform in terms of both radiation amplitude and pulse phase. As the charge value increases, the effective spacetime potential compresses the perigee region of the orbit inward, causing the overall geometric dimensions of the orbit to contract and the inner‑layer whirl structure to become more tightly wound. At $q = 0$, corresponding to the standard charge‑free Kerr spacetime, the radial extent of the orbit reaches its maximum. The peak amplitude of the gravitational wave pulse increases in tandem with the charge, while the radial scale at the periastron continuously decreases as $q$ increases, further amplifying the $1/r$ radiation weighting. The pulse peak corresponding to $q = 0.35$ is significantly higher than that of the Kerr spacetime curve at $q = 0$. Charged spacetime alters the relative rates of orbital radial and azimuthal precession. The appearance times of pulse clusters exhibit slight time shifts for different charge values, and the pulse curves from multiple sets cannot be perfectly aligned. The magnitude of the black hole’s charge can be distinguished based on the characteristic phase shifts in the waveforms. The low‑frequency, slow envelope evolution trends of the waveforms remain consistent for all charge values, with differences limited to pulse details and peak heights. This demonstrates that the orbital resonance period and the macroscopic envelope shape are uniquely determined by the topological parameters $(z, w, v)$, while the NED charge only modifies the radiation intensity (see, e.g., Ref. \cite{48}).

\begin{figure*}[!htp]
\centering
\includegraphics[width=0.95\textwidth]{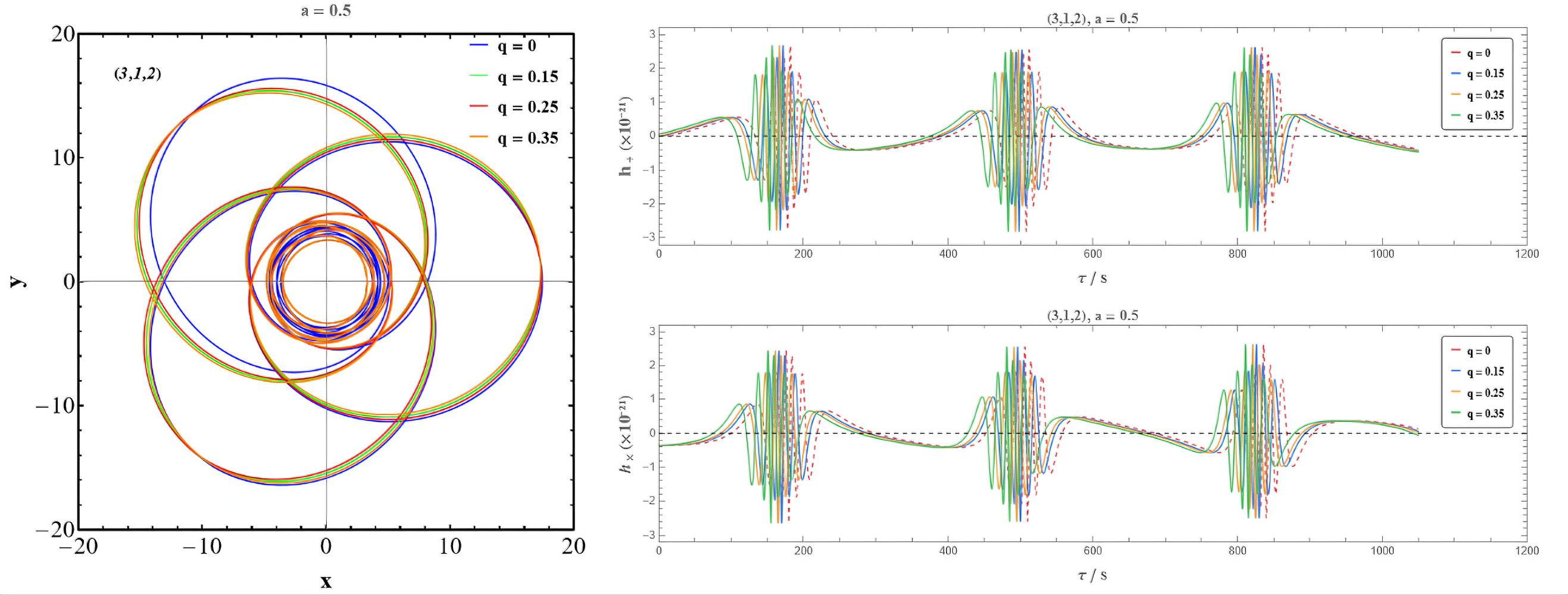}
\caption{Gravitational waveforms corresponding to different charge parameters for the periodic orbit $(3,1,2)$. The images on the left show the periodic orbits for different charge parameters, while the images on the right show the gravitational waveforms radiated by these periodic orbits. The top and bottom images correspond to different polarization directions, respectively.}
\label{fig:fig9}
\end{figure*}

For the fixed resonance topology $(3,1,2)$ with NED charge $q = 0.25$, adjusting the black hole spin parameter $a$ alters the frame‑dragging effect in Kerr spacetime, simultaneously regulating the orbital precession period and the high‑frequency pulse density of the waveform. The higher the spin value, the stronger the spacetime dragging effect. The overall orbital precession period continues to lengthen, and the fluctuations in the low‑frequency envelope of the waveform become smoother. At $a = 0.7$, the rate of envelope evolution is slowest, while at $a = 0.2$ (low spin), the envelope exhibits more pronounced fluctuations and a greater number of low‑frequency oscillations. Under low‑spin conditions, the orbital azimuthal precession rate is faster. Within the same observation duration, more high‑frequency pulses are generated, and the density of waveform oscillations is significantly higher than that of high‑spin orbits. For high‑spin orbits ($a = 0.7$), the time intervals between pulses are wider, and the duration of individual pulse groups is longer. An increase in spin slightly compresses the perigee radial scale of the orbit, and the peak amplitude of the pulses rises gradually as spin increases. At the same time, changes in the spin parameter alter the contribution of the observation inclination coupling term, resulting in subtle distinctions in the amplitude stratification between the $h_+$ and $h_\times$ sets of polarization waveforms. The orbital geometry and waveform evolution characteristics are fully self‑consistent. As spin increases, the outer‑layer vortices expand outward while the inner‑layer vortices tighten, corresponding to a simultaneous lengthening of the gravitational‑wave low‑frequency envelope period.

\begin{figure*}[!htp]
\centering
\includegraphics[width=0.95\textwidth]{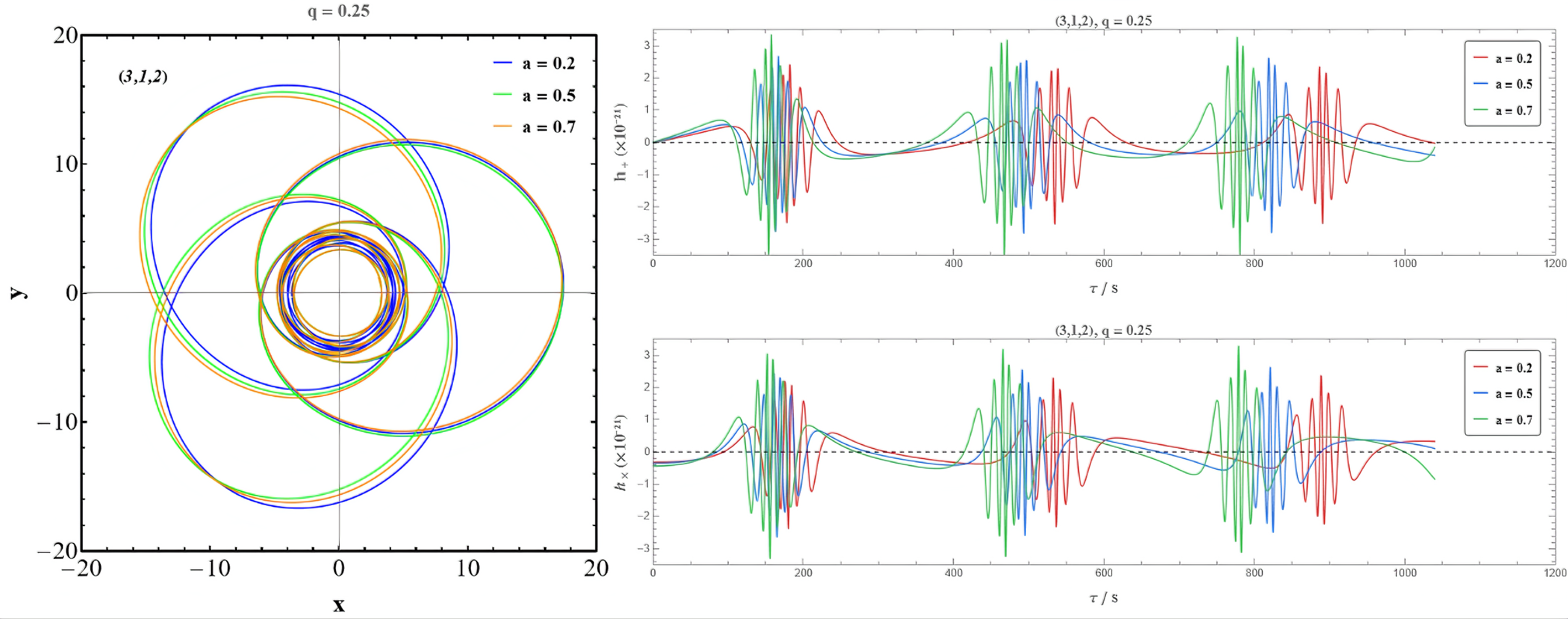}
\caption{Gravitational waveforms for different values of $a$ on the periodic orbit $(3,1,2)$. The images on the left show periodic orbits for different values of $a$, while those on the right show the gravitational waveforms radiated by these periodic orbits. The top and bottom images correspond to different polarization directions, respectively.}
\label{fig:fig10}
\end{figure*}

The resonance topological parameters $(z, w, v)$ are the core physical quantities that determine the macroscopic structure of gravitational wave waveforms. The resonance quality factor $Q$ directly controls the spacing between pulse clusters, the total number of pulses, and the duration of the complete orbital evolution; it is a signature observational feature that distinguishes different periodic resonance orbits. The NED charge parameter $q$ modulates only the peak radiative amplitude of gravitational wave pulses and introduces a slight phase shift in the pulses, without altering the overall periodic profile of the low‑frequency envelope; the black hole spin $a$ simultaneously regulates both the evolution period of the waveform’s low‑frequency envelope and the density of high‑frequency pulse arrangement, serving as a key waveform feature for distinguishing the magnitude of black hole spin at the observational level. The overall evolutionary trends of the two gravitational wave polarization components, $h_+$ and $h_\times$, are completely synchronized, with only a constant phase difference between them, which fully conforms to the fundamental theory of gravitational wave polarization in general relativity. The crossing of all curves at $\tau=0$ in the numerical plots is a numerical artifact of the choice of integration starting point and global initial phase. This numerical convention does not reflect any intrinsic physical property of the emitted gravitational‑wave signal. The visual effect of staggered curves in the literature (see, e.g., Ref. \cite{58}) can be achieved by using the apocenter as the starting point for integration or by assigning independent phase shifts to different orbits.

\section{Conclusion}\label{sec6}

In this work we investigate test‑particle dynamics and gravitational‑wave emission for EMRI systems around rotating charged black holes in the Einstein–Euler–Heisenberg nonlinear electrodynamics model. We sequentially address the construction of spacetime structures and geodesics, the analysis of critical bound orbits and $E$–$L$ phase space, the topological classification of periodic resonant orbits, orbital zoom‑whirl geometric evolution, and modulation of gravitational‑wave polarization waveform parameters, and systematically compare the dynamical differences between the Kerr no‑hair solution and hairy NED black‑hole spacetimes. All research conclusions can be integrated into four major aspects: spacetime modifications, orbital dynamics, periodic‑orbit topology, and gravitational‑wave observational characteristics.

From the perspective of the fundamental structure of spacetime, we obtain the spacetime metric for an $s=2$ rotating charged NED black hole expressed within Boyer‑Lindquist coordinates. We clarify that the NED correction term $\frac{2 q^{4}}{5 r^{4}}$ significantly alters the curvature of spacetime only in the near‑horizon region, while spacetime at large distances degenerates into standard Kerr geometry. By numerically solving the horizon equations, we delineate the range of physical parameters and determine that spin $a$ and charge $q$ work together to lower the upper limit on extreme spins. All selected parameter combinations satisfy the weak energy condition, ensuring the physical consistency of the dynamical calculations. Based on the stationarity and axisymmetry conservation properties of spacetime, we derive the radial effective potential for a test particle on the equatorial plane. The NED charge lowers the effective potential barrier, providing the dynamical basis for the formation of narrow‑band bound orbits near the event horizon and resonant periodic orbits. When $q=0$, the model strictly degenerates into Kerr spacetime, verifying the compatibility of the entire theoretical framework (see, e.g., Refs. \cite{19,43,44}).

Focusing on critical bound orbits and energy‑angular‑momentum coexistence space, we numerically solved the stationary conditions for the marginally‑bound orbit (MBO) and the innermost stable circular orbit (ISCO). Our numerical findings show that black‑hole spin $a$ and NED charge $q$ together yield a combined contracting influence. For both orbit families, radial radius and orbital angular momentum drop in a monotonic fashion when $a$ and $q$ grow larger, while the ISCO binding energy rises simultaneously, making it easier for particles to become unstable and fall into the event horizon. The $E$–$L$ phase space intuitively illustrates the existence boundaries of bound orbits; an increase in charge continuously compresses the phase‑space range and shifts it toward the low‑angular‑momentum region. This compression effect is significantly amplified in high‑spin environments, while the charge‑free Kerr spacetime possesses the widest permissible range of orbital parameters. Relying on the number of zeros of the radial discriminant function, we achieve a complete classification of orbits, clearly distinguishing between absorption orbits, stable bound orbits, MBO critical orbits, and scattering orbits. We establish that the ISCO energy is the lower bound for stable bound orbits, while $E=1$ is the upper bound for orbital escape energy, thereby defining strict physical boundaries for subsequent parameter screening of periodic orbits (see, e.g., Refs. \cite{44,45}).

In the context of periodic resonant‑orbital dynamics, this paper employs a $(z,w,v)$ ternary integer coding system to characterize closed resonant orbits originating from the work of Levin and Perez‑Giz \cite{34}. The resonance ratio $Q=w+\frac{v}{z}$ determines the overall precession characteristics of the orbit. The radial oscillation number $z$ corresponds to the number of orbital lobes, and the zoom number $w$ controls the scale of the pericenter whirl rotation. The orbital precession ratio $Q$ grows monotonically as particle orbital energy becomes larger, yet it falls steadily when orbital angular momentum takes higher values. NED charge suppresses the azimuthal precession rate and reduces the precession ratio at a given energy level, while spin amplifies the modulation distinction of charge on precession characteristics. We numerically construct multiple sets of typical periodic orbits and extract corresponding energy and angular‑momentum data tables. We confirm that increases in both $a$ and $q$ compress the orbital radial activity range, causing the orbit to contract inward as a whole, significantly increasing the degree of whirl entanglement at the periapsis, reducing the proportion of the zoom phase, and making the phase of intense whirl motion the dominant geometric feature of the orbit. The greater the number of orbital lobes $z$, the higher the topological complexity of the orbit; orbits with different topologies exhibit distinguishable geometric projection morphologies (see, e.g., Refs. \cite{47,55}).

Drawing upon the adiabatic approximation alongside the standard quadrupole waveform formalism from Kludge‑based implementations, we calculated the gravitational‑wave polarizations $h_+$ and $h_\times$, establishing a one‑to‑one correspondence between orbital geometry and gravitational‑wave waveforms. The resonant topological parameters $(z,w,v)$ are the core physical quantities that determine the macroscopic low‑frequency envelope of the waveform, the number of pulse clusters, and the total evolution duration. The number of lobes $z$ directly corresponds to the number of oscillation groups in the waveform; orbits with low radial periods have short evolution windows and are prone to numerical integration truncation, whereas orbits with high $z$ have longer evolution periods and exhibit a more uniform and gradual pulse distribution. NED charge acts solely on the peak amplitude and phase shift of pulses; the larger the charge, the smaller the pericentral radial scale. Relying on the $1/r$ radiation amplification effect, it enhances the waveform peak. Different charge curves exhibit distinguishable time shifts but do not alter the overall period of the low‑frequency envelope. Black‑hole spin synchronously modulates the evolution rate of the waveform’s low‑frequency envelope and the high‑frequency pulse density. High spin elongates the envelope period and widens the pulse intervals, while low spin results in a highly fluctuating envelope and a dense pulse arrangement. Spin slightly compresses the radius of the pericentral orbit and simultaneously elevates the pulse amplitude. The evolutionary trends of the two types of gravitational‑wave polarization components are completely synchronized, with only a fixed phase difference, consistent with the theory of gravitational‑wave polarization in general relativity. The fact that waveform curves intersect at zero at the initial moment is solely due to the numerical setting of a unified periastron point and a global phase of zero and does not alter the physical evolution laws of orbital radiation itself (see, e.g., Refs. \cite{57,58}).

From the perspective of theoretical observational value, NED charge and spin leave distinct, layered, and decouplable observational fingerprints in orbital dynamics and gravitational‑wave waveforms. The three categories of parameters — orbital topology, spin, and charge — modulate the macroscopic structure of the waveform, the density of pulses, and the radiation amplitude, respectively. Interplay among these three parameters generates a distinct set of signal features. Such signals enable us to discriminate standard Kerr no‑hair black holes from hairy NED black‑hole geometries. These findings deliver broad theoretical support for next‑generation space gravitational‑wave detectors including LISA, TianQin, and Taiji, which can perform observational tests for the no‑hair theorem and place physical constraints on nonlinear electrodynamics theories (see, e.g., Refs. \cite{48,59}).

\section*{Acknowledgements}

This work was supported by National Natural Science Foundation of China (Grant No. 12365008), the Guizhou Provincial Basic Research Program (Natural Science) (Grant No. QianKeHe‑JiChu[2024]Young166), the Guizhou Provincial Basic Research Program (Natural Science) (Grant No. QianKeHeJiChu‑ZK[2024] YiBan027 and QianKeHeJiChuMS[2025]680), the Guizhou Provincial Major Scientific and Technological Program XKBF (2025) 010 (Hosted by Professor Xu Ning), the Guizhou Provincial Major Science and Technological Program XKGF (2025) 009 (Hosted by Professor Xiang Guoyong) and the Guizhou Provincial Major Scientific and technological Program (Teacher Fan Lu Lu moderated).




\bibliography{ref}
\end{document}